\documentclass[aps,pre,showpacs,english,floats,superscriptaddress,floatfix,notitlepage]{revtex4-1}
\usepackage{bm}
\usepackage{url}
\usepackage{times}
\usepackage{verbatim}
\usepackage{theorem}
\usepackage{makeidx}
\usepackage{amsmath}
\usepackage{siunitx}
\usepackage{xy}
\usepackage{amssymb}
\usepackage{graphicx}
\usepackage{color}
\usepackage{epstopdf}
\usepackage{epsfig}

\usepackage{subfigure}  
\usepackage{hyperref}   

\makeatother

\usepackage{babel}
\begin{document}
\title{Non-Precipitating Shallow Cumulus Clouds: Theory and Direct Numerical Simulation}
\author{S. Ravichandran}
\email{ravichandran@su.se}
\affiliation{Nordita, KTH Royal Institute of Technology and Stockholm University, Stockholm, Sweden, 10691}
\author{Roddam Narasimha}
\affiliation{Engineering Mechanics Unit, Jawaharlal Nehru Centre for Advanced Scientific Research, Bangalore, India 560064}
\email{roddam@jncasr.ac.in}

\begin{abstract}
We model cumulus clouds as transient diabatic plumes, and present
a single-fluid formulation for the study of the dynamics of developing shallow
cumulus clouds. The fluid, air, carries water vapour and liquid water as scalars
that are advected with the flow. We employ the Boussinesq approximation
and a simplified form of the Clausius-Clapeyron law for phase changes.
We use this formulation to study the ambient conditions required for the 
formation of cumulus clouds. We show that the temperature lapse
rate in the ambient, $\Gamma_{0}$; and the relative humidity of the
ambient, $s_{\infty}$, are crucial in deciding when cumulus clouds form.
In the phase-space of these two parameters, we determine the boundary that 
separates the regions where cumulus clouds can form and where they cannot.
\end{abstract}

\maketitle

\section{Introduction\label{sec:Introduction}}

Cumulus clouds, from the Latin for `heap', can be of a wide range
of sizes and take a variety of shapes. The classification of cloud
types by the WMO \citep{wmocloudatlas} recognises several species of cumulus
clouds depending on shape, size and altitude. The common feature uniting
these various types is that cumulus clouds are typically active; that
is, they consist of strong updraughts driven by local sources of buoyancy,
and have lifetimes that last only as long as the updraughts that feed
them. Cumulus lifetimes can vary from less than an hour to a few hours.\\

Clouds are, in general, suspensions of liquid water droplets, and possibly ice particles as well,
of various sizes in a mixture of air and water vapour \citep{wmocloudatlas}.
Clouds play a crucial role in the planet's climate, and are the last
bastion of uncertainty in climate modelling, being responsible for
significant feedback on both short-wave and long-wave radiation 
\citep{stevens2013climate,bony2015clouds}.
The response of this radiative feedback to surface conditions is the focus of climate research.
This response depends on the amount of energy and the mass of water
vapour that is transported from the surface to higher altitudes, and on
the variation of these fluxes with changes in surface conditions.
The transport of thermal energy and water vapour from the surface to higher
altitudes occurs through the formation of clouds in the atmosphere.\\

As the atmosphere generally gets colder and lighter with increasing altitude,
water vapour cools and condenses into liquid water droplets as it rises in 
the atmosphere. This change of phase releases a significant
amount of energy into the flow, changing the nature of the flow. Clouds
are thus not only markers of the patterns of flow in the atmosphere,
but also regions where the flow in the atmosphere is qualitatively
different from its surroundings. Because of the qualitative difference
between ascending flow (with saturated vapour) and descending flow
(with unsaturated vapour) in the atmosphere, convection in the atmosphere
often takes the form of narrow strong upwelling regions (called hot towers)
surrounded by broad regions of weak downward flow. This was first
recognised by \cite{bjerknes1938saturated}. These tall towers of
upward flow are thus responsible for the upward transport of mass
and thermal energy in the atmosphere; these fluxes are controlled by the interaction
of the cloud flow with the ambients. Hot towers are one common form of cumulus clouds,
and their interaction with ambient fluid results in entrainment into or detrainment from
the cloud flow.\\

The evolution of a cumulus cloud depends on surface and ambient conditions,
including the surface temperature, the ambient temperature and its
rate of decrease with height, the ambient aerosol content (amount,
type, size distribution), and covers length scales from fractions 
of a micrometre \citep[Chapter 2]{pruppacher_klett} to 
the characteristic dimensions of the cloud---a few hundred metres
in width, and up to several kilometres in height \citep{RN2011PNAS}.
The amount of liquid water (and solid ice in some cases), and the size
distributions of the particles making up these phases, also vary over
the lifetime of a cumulus cloud; water droplets of a sufficiently
large size (typically about a millimetre; see \citet[Chapter 2]{pruppacher_klett})
fall out of the cloud as precipitation. Given the wide range of lengthscales
and the variety of physical interactions invovled, a complete and accurate description
of the evolution of a cloud is impossible, either through observations
or through computer simulations \citep{grabowskiARFM,randall2003}.
Simplified descriptions of the dynamics are thus required. These simplifications
involve neglecting processes that are unimportant for certain types
of cloud and/or during certain periods of the cloud's lifetime.\\

In this paper, we study the fluid dynamics of a non-precipitating
shallow cumulus cloud, modelled as an isolated turbulent plume from
a finite source of buoyancy. Ignoring precipitation obviates the modelling
of the suspended phase, which may then be treated as a scalar carried
with the flow while retaining the thermodynamics of condensation and
evaporation. Restricting ourselves to shallow clouds allows us to ignore
compressibility effects and employ the Boussinesq equations of motion
(Eq. \ref{eq:Boussinesq_factor}). \\

The problem of entrainment in free-shear flows is of significant fluid
mechanical interest, and has been the object of study for several
decades \citep{MTT1956,Turner1962}. Cumulus clouds, being free-shear
flows, are thus of interest in understanding this fundamental process
in free-shear flows. In addition, as mentioned above, cumulus clouds
and their interaction with their ambients are a crucial and as-yet
poorly understood piece of the global climate puzzle. Most GCMs model
cumulus clouds as steady plumes \citep{derooy2013}. This is inconsistent
with well-known results that heated free-shear flows have anomalous
entrainment (see, e.g., \citet{Bhat1996}). A model for cumulus
clouds incorporating the effects of anomalous entrainment is the transient
diabatic (as opposed to adiabatic) plume, introduced in \cite{RN2011PNAS}
and shown to be able to reproduce commonly observed cloud shapes.
It was also shown in the latter study that the Reynolds number, as
long as it is large enough, is not a crucial parameter, an idea we
will use repeatedly. We note in passing that shallow vigorous cumulus
clouds may be modelled as bubbles/thermals; first introduced by \cite{Turner1969ARFM},
an idea that has become popular again \citep{jeevanjee2019entrainment}.\\

Convection in the atmosphere is also modelled as moist Rayleigh-Benard
convection \citep{pauluis2010idealized,schumacher2010moistrbc,wu2009}. 
We note that highly turbulent Rayleigh-Benard convection takes
the form of penetrative plumes \citep{kadanoff2001turbulent,arakeri2005plume},
which are responsible for significant portions of the energy and mass transfer
in the flow. The approach here studies, without the horizontal movement that 
normally occurs in a Rayleigh-Benard setup, a single plume and its interaction
with its ambient. The modelling of cumulus clouds as isolated
turbulent plumes is thus complemetary to the study of moist Rayleigh-Benard
convection. \\

The rest of the paper is organised as follows. In \S \ref{sec:setup},
the nondimensional governing equations are written down and simplified,
and these simplifications justified. In \S \ref{sec:simulations},
the numerical algorithm used in our study is described. Results from
the simulations are presented in \S \ref{sec:Results}, and
compared with those from a simple one-dimensional model (\S \ref{subsec:one_d})
along the lines of \cite{MTT1956}. We convert to scales typical
of the atmosphere and comment on what the nondimensional solutions
mean for clouds. We conclude in \S \ref{sec:Conclusion}.

\section{Problem Setup\label{sec:setup}}

The dynamics of interest involves the rise of a plume, carrying water
vapour, in a stratified ambient with a known relative humidity. When
the vapour pressure is greater than the local saturation vapour
pressure, it condenses into liquid water, heating the flow and increasing
its buoyancy. A description of the flow thus needs to incorporate
the dynamics of a turbulent plume in a stratified ambient along with
the thermodynamics of phase-change of water, from vapour to liquid
and vice-versa. It will be useful to begin by writing down the conventions
used for naming variables and parameters.

\subsection{Notation \label{subsec:Notation}}

The following conventions are used for all symbols.
\begin{itemize}
\item The vertical direction is $x$, and gravity points along $-\boldsymbol{e}_{x}$.
\item Dimensional quantities are denoted by symbols with tilde ($\tilde{}$);
the tilde is dropped once the variable is nondimensionalised. 
\item Symbols with a subscript infinity $_{\infty}$ denote ambient quantities
(usually functions of the height above ground level) away from, and
not affected by, the flow. 
\item Quantities at ground level are given the subscript zero ($_{0}$). 
\item The water-substance mixing ratios (defined in eq. \ref{eq:mixing_ratio_defn})
are denoted with a hat ($\hat{}$) before they are normalised using
a base value (see eq. \ref{eq:scaled_Clausius_Clapeyron}). 
\item Combinations of the above are possible: the base temperature, for
instance, is $T_{\infty,0}$ (eq. \ref{eq:temp_profile}). 
\end{itemize}

\subsection{The stratified ambient\label{subsec:stratified_plumes}}

\begin{figure}
\noindent \centering{}\includegraphics[width=1\columnwidth]{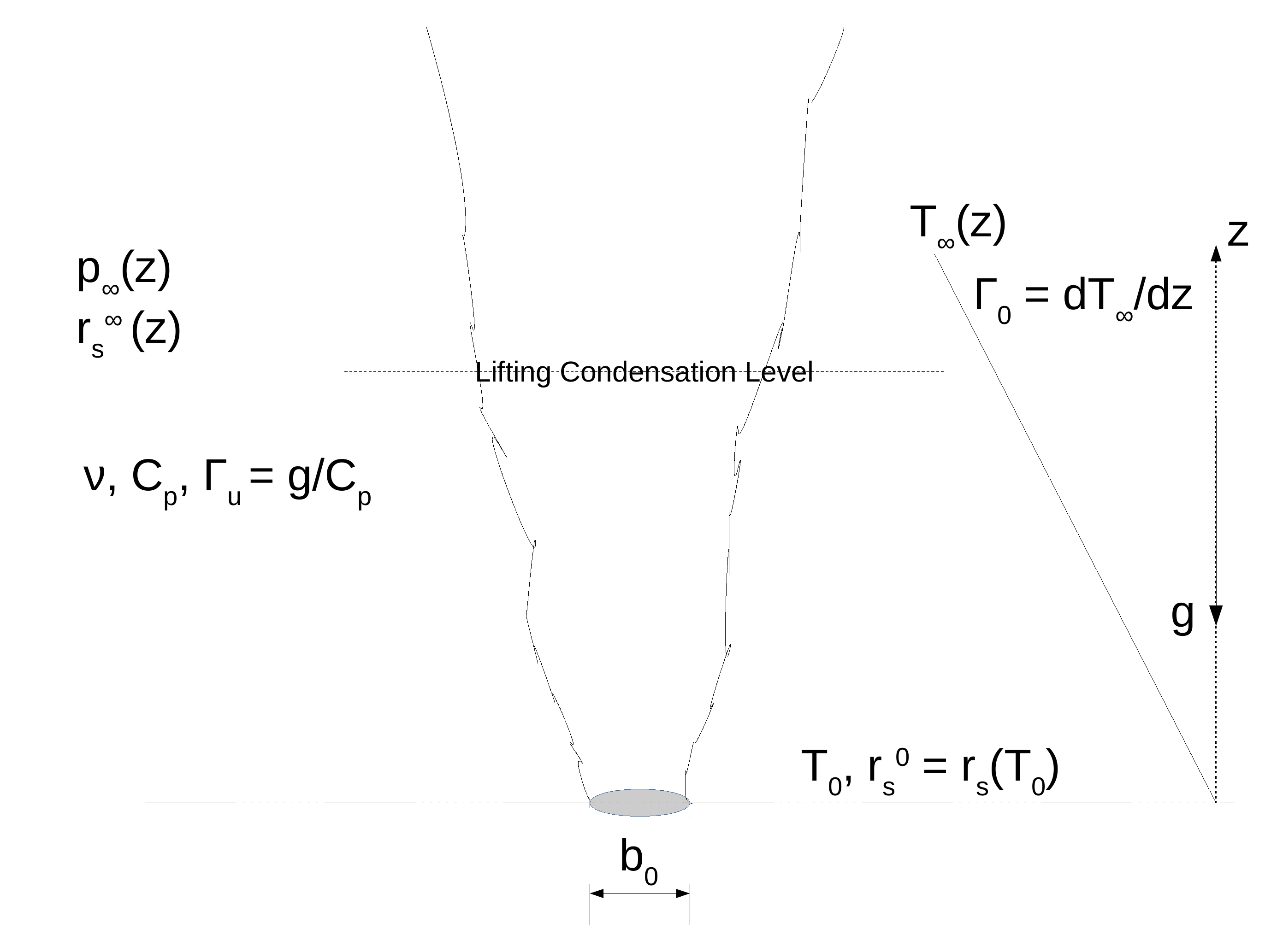}
\caption{Schematic depiction of the system. The hot patch in grey is a source
for both temperature and vapour.}
\label{fig:schematic}
\end{figure}

The plume emanates from a circular patch on the ground of diameter
$b_{0}$, which we will choose as the length-scale. This hot-patch
has a temperature difference $\Delta T_{0}$ over the rest of the
surface at $x=0$, which is at a temperature $T_{\infty,0}$. We will
use $\Delta T_{0}$ as the scale for temperature differences. Together,
$T_{\infty,0}$ and $\Delta T_{0}$ define the Atwood number $\epsilon=\Delta T_{0}/T_{\infty,0}$,
which is typically $\mathcal{O}\left(10^{-2}\right)$. We will use
the fact that this ratio is small in what follows. For the stratification
to be stable, the density of the fluid must decrease with height above
ground level. The strength of the stratification is given in terms
of the rate of decrease of density with height; the larger the rate,
the stabler the stratification.

In a dry gas, density and temperature are connected via the gas law
$p_{\infty}\left(x\right)=\rho_{d,\infty}R_{d}T_{\infty}$, with $p_{\infty}\left(x\right)$
being the ambient pressure at a height $x$ above ground-level, $\rho_{d,\infty}$
the density of dry air, and $R_{d}$ the gas constant for dry air.
The stratification may thus be prescribed in terms the ambient temperature
$T_{\infty}\left(x\right)$ which decreases linearly with height:

\begin{equation}
T_{\infty}\left(x\right)=T_{\infty,0}-\tilde{\Gamma}_{0}x,
\label{eq:temp_profile}
\end{equation}
where $T_{\infty,0}=T_{\infty}\left(0\right)$. For a shallow cloud, the
ambient temperature may be assumed to vary linearly with height; and the lapse
rate, i.e. the rate at which the ambient temperature falls with height,
is 
\[
\tilde{\Gamma}_{\infty}\left(x\right)=-\frac{dT_{\infty}}{dx}=\tilde{\Gamma}_{0}=\text{const}.
\]

A dry atmosphere with no radiative effects composed of a single gas
has a dry (subscript $_{d}$) adiabatic lapse rate $\tilde{\Gamma}_{d}=g/C_{p}$, 
where $g$ is the acceleration due to gravity and $C_{p}$ is the specific heat 
of the gas at constant pressure. The actual lapse rate $\tilde{\Gamma}_{0}$ (which is in
general different from $\tilde{\Gamma}_{d}$) must be considered in
comparison to the dry adiabatic lapse rate. For dry convection,
ambient conditions are stably, neutrally or unstably stratified accordingly
as $\tilde{\Gamma}_{0}\lesseqqgtr\tilde{\Gamma}_{d}$ (see, e.g. \cite{pauluis2010idealized}).
The lapse rates $\tilde{\Gamma}_{d}$ and $\tilde{\Gamma}_{0}$ are
nondimensionalised using $\Delta T_{0}$ and $b_{0}$:

\begin{align*}
\Gamma_{d} & =\frac{\tilde{\Gamma}_{d}b_{0}}{\Delta T_{0}}=\frac{\tilde{\Gamma}_{d}b_{0}}{T_{\infty,0}\epsilon},\\
\Gamma_{0} & =\frac{\tilde{\Gamma}_{0}b_{0}}{\Delta T_{0}}=\frac{\tilde{\Gamma}_{0}b_{0}}{T_{\infty,0}\epsilon}.
\end{align*}
We will report our results in terms of the above nondimensional lapse-rates.

The pressure $p_{\infty}\left(x\right)$ in the ambient is given by

\begin{equation}
\frac{p_{\infty}\left(x\right)}{p_{\infty,0}}=\left(\frac{T_{\infty}\left(x\right)}{T_{\infty,0}}\right)^{g / \tilde{\Gamma}_0 R_d}.\label{eq:press_profile}
\end{equation}
In making the Boussinesq approximation for shallow clouds (discussed in section
\ref{subsec:Boussinesq}), we set this ratio to unity, and therefore
assume that $p_{\infty}\left(x\right)=p_{\infty,0}$.

The buoyancy $B$ of a cloud parcel is given by the density difference
between the parcel and the ambient. The density of a parcel containing both vapour and liquid
water components is 
\[
\rho=\rho_{d}\left(1+\tilde{r}_{v}+\tilde{r}_{l}\right),
\]
which is a reasonable assumption when $\tilde{r}_{v}, \tilde{r}_{l} \ll 1$. The
density of the ambient, with the background mixing ratio $\tilde{r}_{v,\infty}$
of vapour, is similarly given by
\[
\rho_{\infty}=\rho_{d,\infty}\left(1+\tilde{r}_{v,\infty}\right).
\]
In open flows such as we have, the thermodynamic pressure inside the
cloud can be assumed to be the same as the ambient pressure. We thus
have for the ambient 
\[
p_{\infty}=p_{d,\infty}+p_{v,\infty}=\rho_{d,\infty}\left(1+\frac{R_{v}}{R_{d}}\tilde{r}_{v,\infty}\right)R_{d}T_{\infty}=\rho_{d,\infty}\left(1+\chi\tilde{r}_{v,\infty}\right)R_{d}T_{\infty},
\]
where $\chi$ is the ratio of the gas constants of air and water,
$\chi=R_{v}/R_{d}=28.9/18=1.61.$ Similarly, for a cloud parcel,

\begin{align*}
p_{\infty} & =\rho_{d}R_{d}\left(1+\frac{R_{v}}{R_{d}}\tilde{r}_{v}\right)T.
\end{align*}
Thus, the ratio of densities is

\begin{align*}
\frac{\rho}{\rho_{\infty}} & =\frac{\rho_{d}\left(1+\tilde{r}_{v}+\tilde{r}_{l}\right)}{\rho_{d,\infty}\left(1+\tilde{r}_{v,\infty}\right)}=\left(\frac{T_{\infty}}{T}\right)\left(\frac{1+\chi\tilde{r}_{v,\infty}}{1+\chi\tilde{r}_{v}}\right)\left(\frac{1+\tilde{r}_{v}+\tilde{r}_{l}}{1+\tilde{r}_{v,\infty}}\right)\\
 & =\frac{T_{\infty}}{T}\left[1+\left(\chi-1\right)\left(\tilde{r}_{v,\infty}-\tilde{r}_{v}\right)+\tilde{r}_{l}\right]\\
 & =\frac{T_{\infty}}{T}+\frac{T_{\infty}}{T}\left[\left(\chi-1\right)\left(\tilde{r}_{v,\infty}-\tilde{r}_{v}\right)+\tilde{r}_{l}\right].
\end{align*}
It follows that
\begin{align*}
1-\frac{\rho}{\rho_{\infty}} & =1-\frac{T_{\infty}}{T}-\frac{T_{\infty}}{T}\left[\left(\chi-1\right)\left(\tilde{r}_{v,\infty}-\tilde{r}_{v}\right)+\tilde{r}_{l}\right]\\
 & \approx\frac{T-T_{\infty}}{T_{\infty}}-\left(\left(\chi-1\right)\left(\tilde{r}_{v,\infty}-\tilde{r}_{v}\right)+\tilde{r}_{l}\right),
\end{align*}
where we have taken $T/T_{\infty}\approx1$ in the second term and
replaced the denominator $T$ in the first term with $T_{\infty}$.
We then have, for the buoyancy,

\begin{equation}
B=g\frac{\rho_{\infty}-\rho}{\rho_{\infty}}=g\left[\frac{T-T_{\infty}}{T}+\left(\chi-1\right)\left(\tilde{r}_{v}-\tilde{r}_{v,\infty}\right)-\tilde{r}_{l}\right].\label{eq:buoyancy_dim}
\end{equation}
If we assume that temperature differences are $\mathcal{O}\left(\Delta T\right)$,
the buoyancy $B=\mathcal{O}\left(g\epsilon\right)$. The velocity
scale in the problem has to be derived from the \emph{a priori} assumption
that the buoyancy forces in the problem are $\mathcal{O}\left(1\right)$
compared to the advective forces. That is,

\begin{equation}
Fr^{2}=\frac{W^{2}/b_{0}}{g\epsilon}=\mathcal{O}(1),
\end{equation}
giving, for the velocity scale, 
\begin{equation}
W=\sqrt{b_{0}g\epsilon}.
\end{equation}
We work with absolute temperatures (instead of potential temperatures
as is more common in the atmospheric sciences). Temperature differences
are scaled in units of $\Delta T_{0}$, giving the nondimensional
temperature differential.

\begin{equation}
\theta=\frac{T-T_{\infty}\left(x\right)}{\Delta T_{0}}.
\end{equation}
We will assume that $\nu$ the kinematic viscosity of the fluid is
also known and constant. Together with the length and velocity scales
defined above, this defines the Reynolds number 
\begin{equation}
Re=\frac{Wb_{0}}{\nu}=\frac{b_{0}\sqrt{b_{0}g\epsilon}}{\nu}.
\end{equation}
The Prandtl number $Pr=\nu/\kappa$, the ratio between the diffusivities
of momentum and temperature, is $\mathcal{O}\left(1\right)$ for gases.
For simplicity, we take $Pr=1$.

\subsection{Thermodynamics\label{subsec:thermodynamics}}

Cloud flows are complicated by the addition of the 
thermodynamics of phase-change to the dynamics of plumes in 
stratified ambients. The amounts of water vapour and liquid 
water present are given in terms of the mixing ratios of 
vapour (subscript $_{v}$) and liquid (subscript $_{l}$) 
respectively. These are defined as the mass of water substance
per unit mass of dry air. For instance, the mixing ratio of vapour
is 
\begin{equation}
\hat{r}_{v}=\frac{\rho_{v}}{\rho_{d}}.\label{eq:mixing_ratio_defn}
\end{equation}
The thermodynamics of water is governed by the Clausius-Clapeyron
law which gives the saturation mixing ratio of vapour at a temperature
$T$ and pressure $p_{\infty}\left(x\right)$, and can be written
as

\begin{equation}
\hat{r}_{s}\left(T,p_{\infty}\left(x\right)\right)=r_{s,0}\frac{p_{\infty,0}}{p_{\infty}\left(x\right)}\mbox{exp}\left(\frac{L_{v}}{R_{V}}\left(\frac{1}{T_{\infty,0}}-\frac{1}{T}\right)\right),\label{eq:Clausius-Clapeyron}
\end{equation}
where $r_{s,0}$ is the saturation mixing ratio at the ground-level
temperature and pressure $\left(T_{\infty,0},p_{\infty,0}\right)$.
In addition, $L_{v}$ is the latent heat of vapourisation of water,
and $R_{v}$ is the gas constant for water vapour. We assume that
the ambient has a background concentration of water vapour, given
by 
\begin{equation}
\hat{r}_{v}^{\infty}=s_{\infty}\left(x\right)\times\hat{r}_{s}\left(T_{\infty}(x),p_{\infty}(x)\right),\label{eq:background_vapour}
\end{equation}
with $s_{\infty}\left(x\right)$ being the relative humidity of the
ambient which can be a function of altitude. The corresponding background
concentration of liquid water is $\hat{r}_{l}^{\infty}=0$.

We will make a few simplifying assumptions. First, in defining the
mixing ratio $\hat{r}_{l}$ for liquid water in the flow, we ignore
the details of the size-distribution of the water droplets that make
up the liquid water. Called the one-moment scheme in atmospheric physics,
this is justifiable for non-precipitating clouds (see, e.g. \cite{morrison2009}).
We will also assume that the liquid field thus defined is advected
with the flow, also justifiable if the liquid content is made up of
a very large number of small droplets. This is again true for non-precipitating
clouds, in which the size-distribution of the droplets peaks at around
$10\mu$m.

We will also, henceforth, scale all mixing ratios in units of the
saturation mixing ratio $r_{s,0}\left(T_{\infty,0},p_{\infty,0}\right)$
at the ground-level, and denote these scaled mixing ratios by dropping
the $\hat{}$. Thus, in these scaled units, eq. \ref{eq:Clausius-Clapeyron}
becomes

\begin{equation}
r_{s}\left(T,p_{\infty}\left(x\right)\right)=\frac{\mbox{exp}\left(\frac{L_{v}}{R_{V}}\left(\frac{1}{T_{\infty,0}}-\frac{1}{T}\right)\right)}{p_{\infty}\left(x\right)/p_{\infty,0}}.\label{eq:scaled_Clausius_Clapeyron}
\end{equation}
In addition, we assume that the species Prandtl numbers associated
with vapour and liquid diffusivities, denoted by $Pr_{v}$ and $Pr_{l}$
respectively, are both unity. This is a reasonable assumption for
the vapour. Liquid droplets in evolving cumulus clouds have a size distribution
that peaks around $10\mu$m \citep{pruppacher_klett}, and are therefore not influenced by Brownian diffusion. As a first approximation, however, we proceed with this assumption.
Small-scale features of the flow will, as a result, be affected.

\subsubsection{Supersaturation and thermodynamics of phase-change}

We assume, further, that mixtures of air and water vapour can be supersaturated;
i.e. that more vapour than prescribed by the Clausius-Clapeyron law
(eq. \ref{eq:Clausius-Clapeyron}) can exist in a parcel of a given
temperature. In such supersaturated parcels, the water vapour condenses
into liquid water at a rate which depends on the supersaturation:

\[
\frac{dr_{v}}{dt}\sim\left(\frac{r_{v}}{r_{s}}-1\right),
\]
where the term on the right hand side is called the supersaturation.
The supersaturation in atmospheric clouds is typically of the order
of a few percent. The constant of proportionality in this equation
is a timescale $\tilde{\tau}_{s}$, which can be derived by assuming
the droplet number and distribution (see, e.g., \cite{KumarShaw2014}).
This timescale is strictly a function of time and space, and its variation
can affect the flow. However, if the droplet size distribution is
monodisperse, and we consider only non-precipitating clouds, $\tau_{s}$
is typically small compared to the timescale $b_{0}/W$ of the flow.
In this work, we choose a small value for 
\[
\tau_{s}=\frac{\tilde{\tau}_{s}W}{b_{0}}=0.1,
\]
and do not discuss the effects of its variations. Thus, the rate of
phase-change (in nondimensional variables) is given by

\begin{equation}
\frac{dr_{l}}{dt}=-\frac{dr_{v}}{dt}=\mathcal{H}\frac{1}{\tau_{s}}\left(\frac{r_{v}}{r_{s}}-1\right),\label{eq:thermo}
\end{equation}
where $\mathcal{H}$ is a modified heaviside function, which is positive
if the parcel is supersaturated, or if the parcel is subsaturated
but has nonzero liquid content:

\[
\mathcal{H}=\begin{cases}
1 & \text{if }r_{v}>r_{s}\\
1 & \text{if }r_{v}<r_{s}\text{ and }r_{l}>0\\
0 & \text{if }r_{v}<r_{s}\text{ and }r_{l}=0.
\end{cases}
\]

\subsection{Boussinesq Approximation \label{subsec:Boussinesq}}

\subsubsection{Dynamics}

The Boussinesq approximation involves assuming that flow is incompressible,
even if the ambient is stratified \citep{Spiegel1960}. The incompressibility
condition is the shallow-convection limit of anelastic schemes that
filter out the acoustic modes, so that the bound on the timestep to
be used for numerical integration is relaxed considerably (see, e.g.
\cite{bannon1996anelastic,durran1989improving}).

The Boussinesq approximation is only valid for clouds whose height
is small compared to the density scale height 
\[
H_{\rho}=\left(\frac{1}{\rho}\frac{d\rho}{dx}\right)^{-1}
\]
of the atmosphere. This height is typically $\mathcal{O}\left(10-15\text{ km}\right)$
\citep{bannon1996anelastic}. We are therefore restricted to clouds
which do not rise more than a few kilometres (see, e.g., \cite{bannon1996anelastic,pauluis2010idealized}).

\subsubsection{Thermodynamics}

In order to be consistent in our approximation, we also write the
Clausius-Clapeyron law (eq. \ref{eq:Clausius-Clapeyron}) using the
Boussinesq approximation. This has the added benefit of letting us
describe the problem independently of the absolute temperature $T_{\infty,0}$. First, we will
find it convenient to define the nondimensional constants 
\begin{align*}
L_{1} & =\frac{L_{v}\epsilon}{R_{v}T_{\infty,0}}\text{ and}\\
L_{2} & =\frac{L_{v}r_{s,0}}{C_{p}T_{\infty,0}\epsilon}.
\end{align*}
The first of these quantities appears in the Clausius-Clapeyron equation
\ref{eq:Clausius-Clapeyron} above. The second appears in the temperature
equation. The two quantities are proportional to each other: 
\[
L_{1}=\left(\frac{C_{p}}{R_{v}}\right)\cdot\left(\frac{\epsilon}{r_{s,0}}\right)\cdot L_{2}.
\]
Using $L_{1}$, we rewrite the Clausius-Clapeyron equation as follows.
We take

\begin{align*}
\frac{1}{T_{\infty,0}}-\frac{1}{T} & =\frac{\Delta T_{0}\theta-\tilde{\Gamma}_{0}\tilde{x}}{T_{\infty,0}\left(T_{\infty,0}-\Gamma_{0}x+\Delta T_{0}\theta\right)}\\
& =\frac{1}{T_{\infty,0}}\left(\frac{\Delta T_{0}\theta-\tilde{\Gamma}_{0}\tilde{x}}{T_{\infty,0}-\Delta T_{0}\Gamma_{0}x+\Delta T_{0}\theta}\right)\\
 & = \frac{\epsilon}{T_{\infty,0}}\left(\frac{\theta-\Gamma_{0}x}{1-\epsilon\left(\Gamma_{0}x+\theta\right)}\right) \approx \frac{\epsilon}{T_{\infty,0}} \left( \frac{\theta-\Gamma_{0}x}{1-\epsilon\Gamma_{0}x} \right)\\
 & =\epsilon \frac{C(x)}{T_{\infty,0}}\left(\theta-\Gamma_{0}x\right),
\end{align*}
where 
\begin{equation}
C(x)=\frac{1}{1-\epsilon\Gamma_{0}x}=\frac{T_{\infty,0}}{T_{\infty}\left(x\right)}>1\label{eq:Boussinesq_factor}
\end{equation}
is a nondimensional factor which is a function of the altitude. In
the Boussinesq approximation, we set $C=1$ and the Clausius-Clapeyron
law (Eq. \ref{eq:scaled_Clausius_Clapeyron}) becomes
\begin{equation}
r_{s}=\text{exp}\left(-L_{1}\Gamma_{0}x\right)\text{exp}\left(L_{1}\theta\right)\label{eq:CC-modified}
\end{equation}
The above equation does not depend directly on $T_{\infty,0}$; as a result, the results here can be directly applied to an arbitrary fluid as long as the correct values for $L_1$ and $L_2$ are specified.

The buoyancy term in the momentum equation has to be carefully approximated
as well. As in \S \ref{subsec:stratified_plumes}, the contribution
from temperature differences is 
\[
B\propto g\frac{T-T_{\infty}\left(x\right)}{T_{\infty}\left(x\right)},
\]
which can be written in our nondimensional variables as 
\[
B\propto\frac{1}{Fr^{2}}\left(C(x)\cdot\theta\right),
\]
where the approximation 
factor $C(x)=1/\left(1-\epsilon\Gamma_{0}x\right)$ appears
again. As before, under the Boussinesq approximation, $C=1$, and
the total buoyancy is

\begin{equation}
B=\frac{1}{Fr^{2}}\left(\theta+r_{0}\left(\left(\chi-1\right)\left(r_{v}-r_{v}^{\infty}\right)-r_{l}\right)\right),\label{eq:buoyancy}
\end{equation}
(see, e.g. \cite{hernandez2013minimal,KumarShaw2014}).

\subsubsection{Validity of the Boussinesq approximation}

\begin{figure}
\noindent \begin{centering}
\includegraphics[width=0.6\columnwidth]{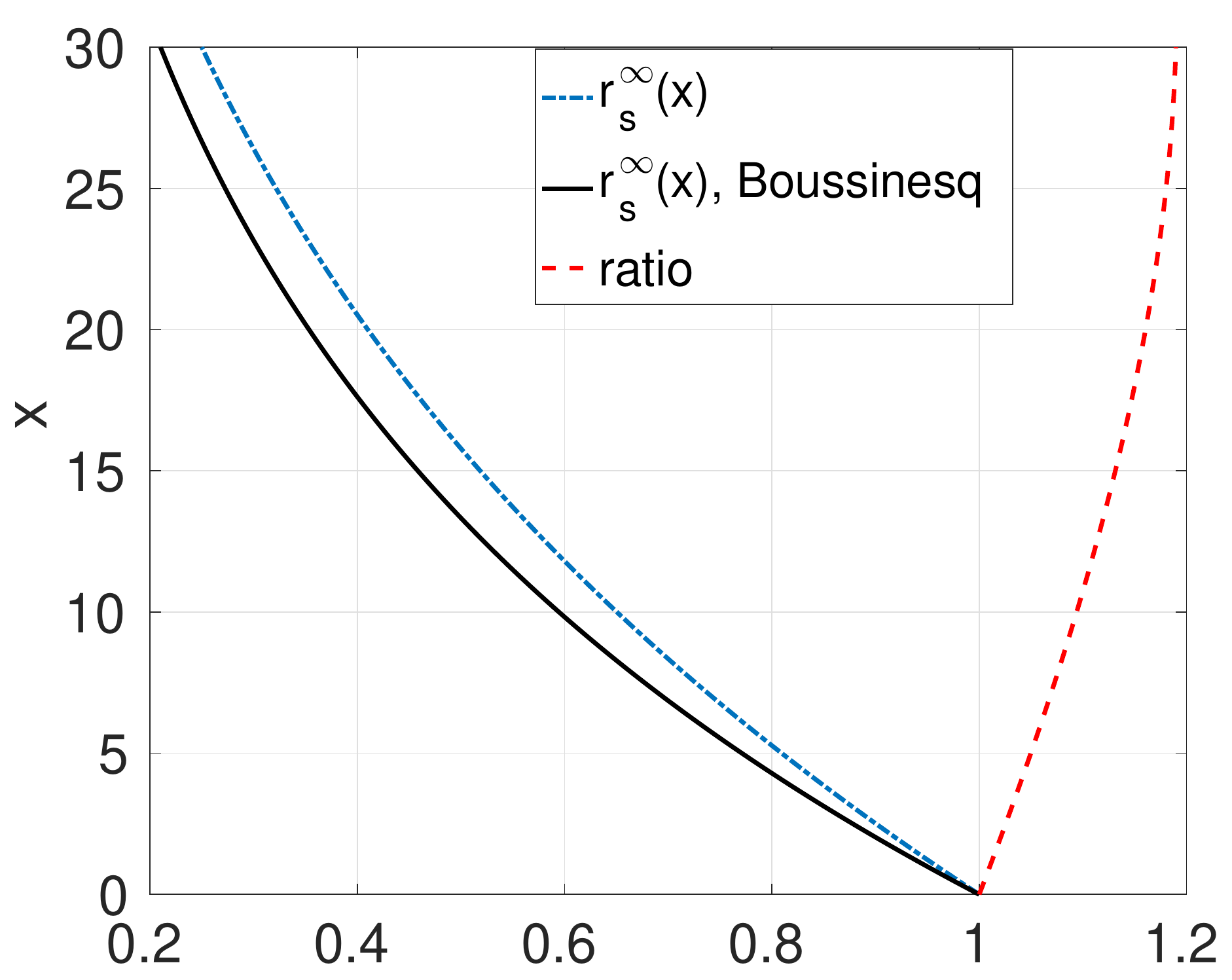}
\par\end{centering}
\caption{\label{fig:rs_inf_vs_x} Validity of the Boussinesq approximation
$C(x)=1$.}
\end{figure}

As mentioned above, the factor $C(x)=\left(1-\epsilon\Gamma_{0}x\right)^{-1}>1$
for a lapsing atmosphere, and the Boussinesq approximation amounts
to setting $C=1$. If the maximum altitude is not too large---i.e.
if the cloud is shallow---the approximation $C\approx1$ is a good
one. The approximation becomes worse with increasing height. This
can be seen in Figure \ref{fig:rs_inf_vs_x}, where the saturation
mixing ratio $r_{s,\infty}\left(x\right)$ is plotted with and without
the Boussinesq approximation $C(x)=1$. The ratio of the two is also plotted. 
The approximation $C=1$ occurs in two places:
in the buoyancy term and in the saturation vapour mixing ratio term.
In the buoyancy term, using the approximation $C=1$ results in an
underestimation of the buoyancy. In the saturation mixing ratio term,
setting $C=1$ underestimates the saturation mixing ratio, and thus
overestimates whether a given mixing ratio of vapour at a given temperature
is saturated.

The maximum value of the factor $C(x)$ occurs when $\Gamma_{0}$
is highest, and at the maximum $x$ of the computational domain. The
maximum value of $\Gamma_{0}$ in our simulations is $\Gamma_{0}=0.09$.
For this case, the maximum value of $C(x)=(1-0.09)^{-1}\approx1.1$.

\subsection{Nondimensional equations and boundary conditions\label{subsec:nondim_eqns}}

The continuity equation, in the Boussinesq approximation, is just
the requirement that the divergence of the velocity be zero:

\begin{equation}
\nabla\cdot\mathbf{u}=0.\label{eq:continuity}
\end{equation}
In writing Eq. \ref{eq:continuity}, we are neglecting the changes in volume
introduced by the condensation of water vapour. This is an approximation, valid
only in the limit as $r_{v,l} \ll 1$.

The momentum equation has contributions from the buoyancy term discussed
in \S \ref{subsec:Boussinesq}, and reads

\begin{equation}
\frac{D\mathbf{u}}{Dt}=-\frac{\nabla p}{\rho_{0}}+\frac{1}{Fr^{2}}\left[\theta+r^{0}\left(\left(\chi-1\right)\left(r_{v}-r_{v}^{\infty}\right)-r_{l}\right)\right]\mathbf{e}_{x}+\frac{1}{Re}\nabla^{2}\mathbf{u},\label{eq:momentum}
\end{equation}
where $D/Dt$ is the material derivative. The factor $L_{2}$ appears
in the temperature equation as follows

\begin{align*}
\frac{D\theta}{Dt} & =\left(\frac{\tilde{\Gamma}_{0}-\tilde{\Gamma}_{d}}{\Delta T_{0}}\right)\tilde{w}+\mathcal{H}\left(\frac{L_{v}r_{s,0}}{C_{p}\Delta T_{0}}\right)\left(\frac{1}{\tilde{\tau}_{s}}\right)\left(\frac{r_{v}}{r_{s}}-1\right)+\frac{1}{Re\cdot Pr}\nabla^{2}\theta,
\end{align*}
which can be nondimensionalised to

\begin{equation}
\frac{D\theta}{Dt}=\left(\Gamma_{0}-\Gamma_{d}\right)w+L_{2}\left[\frac{\mathcal{H}}{\tau_{s}}\left(\frac{r_{v}}{r_{s}}-1\right)\right]+\frac{1}{Re\cdot Pr}\nabla^{2}\theta.\label{eq:theta}
\end{equation}
The vapour and liquid transport equations include the phase-change
terms (eq. \ref{eq:thermo}) and the diffusion terms discussed in
\S \ref{subsec:thermodynamics}:

\begin{equation}
\frac{Dr_{v}}{Dt}=-\left[\frac{\mathcal{H}}{\tau_{s}}\left(\frac{r_{v}}{r_{s}}-1\right)\right]+\frac{1}{Re\cdot Pr_{v}}\nabla^{2}r_{v},\label{eq:vapour}
\end{equation}

\begin{equation}
\frac{Dr_{l}}{Dt}=\left[\frac{\mathcal{H}}{\tau_{s}}\left(\frac{r_{v}}{r_{s}}-1\right)\right]+\frac{1}{Re\cdot Pr_{l}}\nabla^{2}r_{l}.\label{eq:liquid}
\end{equation}

These equations are similar to the non-equilibrium, non-precipitating
equations proposed as a `minimal model' for moist convection by Hernandez-Duenas
\emph{et al}. (2012). The major differences between their (non-equilibrium,
non-precipitating) model and ours are that (a) they use the potential
temperature since they deal with altitudes of up to $15$ km; and
(b) they simplify the thermodynamics such that the saturation vapour
pressure is only a function of the altitude above ground; (c) that
their ambients are always nearly saturated. The first of these assumptions
(i.e. the use of the potential temperature) may be relaxed if convection
is shallow, as described in \S \ref{subsec:Boundary-Conditions}
above. Study of the details of the flow of vapour and liquid water,
as we aim to do, requires local variations of saturation vapour pressure
to be accounted for. Lastly, we study the effects of the extent of
subsaturation on cloud formation.

\subsubsection{Boundary Conditions\label{subsec:Boundary-Conditions}}

We use no-slip wall boundary conditions for the velocity on the bottom
boundary. We impose the no-flux condition on the scalars on the bottom
wall, except for the hot patch, where we prescribe Dirichlet boundary
conditions (described below in more detail). For the velocity components
and the scalars, we impose the Orlanski inflow/outflow boundary conditions
at the five other boundaries of the domain \citep{orlanski1976simple}.
These boundary conditions allow fluid to be entrained or detrained from the sides
of the domain, and flow structures are passively advected at the top
of the box without spurious reflection. Neumann boundary conditions
are applied on the pressure at all boundaries.\\

At the hot patch on the ground, the nondimensional temperature $\theta=1$. We report
results with three boundary conditions for the vapour $r_{v}$ at the hot patch,
which determine the amount of vapour introduced into the flow at the hot patch.
In the first kind, $r_v$ at the hot patch is the ambient value $s_\infty r_s^0$, and
the hot patch introduces no vapour to the system. In the second kind, $r_{v}$ is set to the saturated value at the temperature of the hot patch, viz $r_{v}=r_{s}\left(x=0; \theta=1\right)$.
In the third kind, $r_{v}$ at the hot patch is set to the saturation value at the base temperature $r_s\left(x=0; \theta=0\right)$. These boundary
conditions lead to qualitatively similar results, as we shall discuss in
\S \ref{subsec:vapour_bcs}. Unless otherwise mentioned, we use the saturated
boundary condition for the vapour.

\subsection{Parameters\label{subsec:Parameters}}

Most of the parameters in the problem depend on the Atwood number,
$\epsilon=\Delta T_{0}/T_{\infty,0}$ and the plume-width at base,
$b_{0}$. Given these, the following parameter values are obtained.

\begin{equation}
Re=\frac{Wb_{0}}{\nu}=\frac{b_{0}\sqrt{b_{0}g\epsilon}}{\nu}=\mathcal{O}(10^{7}-10^{8})\gg1
\end{equation}

\begin{equation}
Fr^{2}\equiv1.
\end{equation}
The thermodynamic parameters involve the Atwood number $\epsilon=\Delta T_{0}/T_{\infty,0}$ and the base temperature $T_{\infty,0}$. The problem can be specified independently of  $T_{\infty,0}$ and indeed of the specific combination of fluids. The saturation mixing ratio $r_{s,0}$ is
\begin{equation}
r_{s,0}=\frac{\rho_{s}^{0}}{\rho_{0}}=\mathcal{O}\left(10^{-2}\right)
\end{equation}
for water vapour and air under terrestrial conditions. The other parameters are

\begin{align}
L_1 = \frac{L_{v}r_{s,0}}{C_{p}\Delta T} & =\frac{L_{v}r_{s,0}}{C_{p}T_{\infty,0}\epsilon}\\
\Gamma_{0}=\frac{\tilde{\Gamma}_{0}b_{0}}{\Delta T_{0}} & =\frac{\tilde{\Gamma}_{0}b_{0}}{T_{\infty,0}\epsilon}\\
\Gamma_{d}=\frac{\tilde{\Gamma}_{d}b_{0}}{\Delta T_{0}} & =\frac{g}{C_{p}}\frac{b_{0}}{T_{\infty,0}\epsilon}.
\end{align}
The following values are chosen for all results reported here.
\begin{align*}
\epsilon & = \frac{1}{30}\\
\Gamma_{d,0} & =\frac{9.8\text{ ms}^{-2}\times100\text{ m}}{10^{3}\text{J/kg/K }\times300\text{ K }\times1/30}=0.098;\\
\frac{L_{v}}{C_{p}\Delta T_{0}} & =240\\
r_{s,0} & = 0.015 \\
L_{2}=\frac{L_{v}r_{s,0}}{C_{p}\Delta T_{0}} & \approx3.6\\
L_{1}=\frac{L_{v}\epsilon}{R_{v}T_{\infty,0}} & =0.577.
\end{align*}
We will also need to select the background distributions $r_{v}^{\infty}$ and
$\theta^{\infty}$. The parameters in the problem are
all listed in table \ref{tab:params}.

\begin{table}
\noindent \begin{centering}
\begin{tabular}{|c|l|}
\hline 
Parameter & Type\tabularnewline
\hline 
Re, Pr$=1$ & Flow parameters\tabularnewline
\hline 
$\Gamma_{u}$, $\Gamma_d$, $s_\infty$, $\theta_\infty = 0$, $r_{s,\infty}$, $r_{v,\infty}=s_\infty r_{s,\infty}$ & Ambient stratification properties, assumed constant\tabularnewline
\hline 
$L_1$, $L_2$, $\tau_s=0.1$ & Thermodynamic properties, assumed constant\tabularnewline
\hline
\end{tabular}
\par\end{centering}
\caption{List of nondimensional parameters\label{tab:params}}
\end{table}

\section{Numerical Simulations \label{sec:simulations}}

We solve equations (\ref{eq:continuity}-\ref{eq:liquid}) numerically.
The solver used, dubbed \emph{Megha}-5, the Sanskrit word for cloud, is
currently in its fifth generation. \emph{Megha-5} is a finite volume
solver with a staggered grid---the scalars and the pressure are stored
at cell-centres and the velocity components stored on the cell-faces.
Space-derivatives are calculated using second-order central differences;
timestepping is done using a second-order Adams-Bashforth scheme.
The momentum equation is solved using the operator-splitting method
and the resulting pressure-Poisson equation (PPE) is solved using
three-dimensional discrete cosine transforms (DCTs). We use a modified
wavenumber so that the DCT-based Poisson-solution is also second-order
accurate (see, e.g. \citet{AFiD}). 

Unless otherwise stated, we perform our simulations in a box of size
$15b_{0}\times15b_{0}\times30b_{0}$.

\subsection{Validation\label{subsec:Validation}}

\emph{Megha}-5 is based on an extensively validated earlier version
\citep{PrasanthMS}. The differences are in the incorporation of the
thermodynamics of phase-change, the boundary conditions and the Poisson
solver. \emph{Megha}-5 uses inflow/outflow boundary conditions as
mentioned in \S \ref{subsec:nondim_eqns}. These allow the boundaries
of the domain to be significantly closer to the axis of the plume
than the no-slip walls that were earlier used. The inflow/outflow
boundary conditions have been checked by ensuring that changing the
size of the domain does not significantly alter the solution. The
Poisson solver for the pressure equation, which used to be a multigrid-solver
based on the HYPRE library, was found to be slow and to not scale
when the number of processors was increased. We therefore switched
to a fast-Fourier-transform-based Poisson solver with a modified wavenumber
(see, e.g. \cite{AFiD}), which has an algorithm that scales almost
linearly. The Navier-Stokes portion of the code has been checked by
ensuring that it produces the same solution as the known solution
for the lid-driven cavity. The thermodynamics have also been extensively
tested. The validation case of a spherical thermal has been simulated
and verified. The code has also been used in the study of mammatus clouds
\citep{Mammatus2018}.

\subsection{Resolution requirements \label{subsec:Resolution} }

The grid required has to be such that the smallest scales in the flow---the
Kolmogorov scales---are resolved. We cannot, therefore, simulate
flows with cloud-like Reynolds numbers, since the computational resources
required do not exist today. However, we believe that this limitation
is not fatal to our project, since we only need to reach the mixing
transition Reynolds number of about $10,000$. We report here simulations
with $Re=1000$, with ongoing work at higher Reynolds numbers.


\subsection{Initial conditions \label{subsec:Initcond}}

The flow is started with a patch on the ground that is hotter than
its surroundings by $\epsilon T_{\infty,0}$, with either dry-base
or moist-base vapour boundary conditions. All other variables are
initialised to zero.

\subsection{Turbulence generation \label{subsec:Turbulence-generation}}

Our Reynolds numbers are modest at $Re=\mathcal{O}\left(10^{3}\right)$.
At such low Reynolds numbers, the plume that is obtained in our simulations
remains laminar unless it is forced (or `tripped') into becoming
turbulent. We add a small amount of random noise to the temperature in
order to `trip' the flow into turbulence. The amount of noise added and 
the number of nodes over which the noise is added affect how soon the flow
becomes turbulent (see, e.g. \cite{plourde2008direct}. We add $10\%$ broadband
noise to the temperature over the first $10$ nodes = $0.3$ diameters.
The plume that emanates is turbulent and behaves like a classical plume in which
the buoyancy flux is conserved. This is shown in Figure \ref{fig:pure_plume} below.
Also shown in Figure \ref{fig:pure_plume} is the plume width, calculated using the
mass and momentum fluxes as
\begin{equation}
 \langle b \rangle = \frac{(\langle b^2 U \rangle)^2}{\langle b^2 u^2 \rangle}, \label{eq:plume_width}
\end{equation}
where $\langle b^2 U \rangle (x)$ is the total mass flux  and $\langle b^2 u^2 \rangle (x)$ the total momentum flux through a plane at $x$. 

\begin{figure}
\begin{center}
\includegraphics[width=0.8\columnwidth]{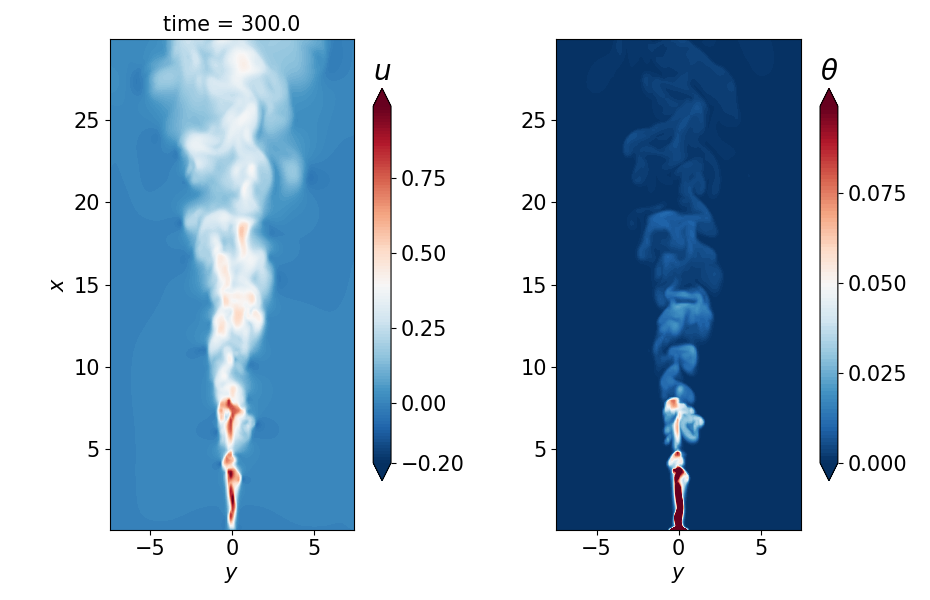}
\end{center}
\caption{\label{fig:pure_plume} (Left) the vertical velocity and (right) the temperature
distributions in a plume at $Re=1000$.}
\end{figure}

\begin{figure}
\begin{center}
\includegraphics[width=0.8\columnwidth]{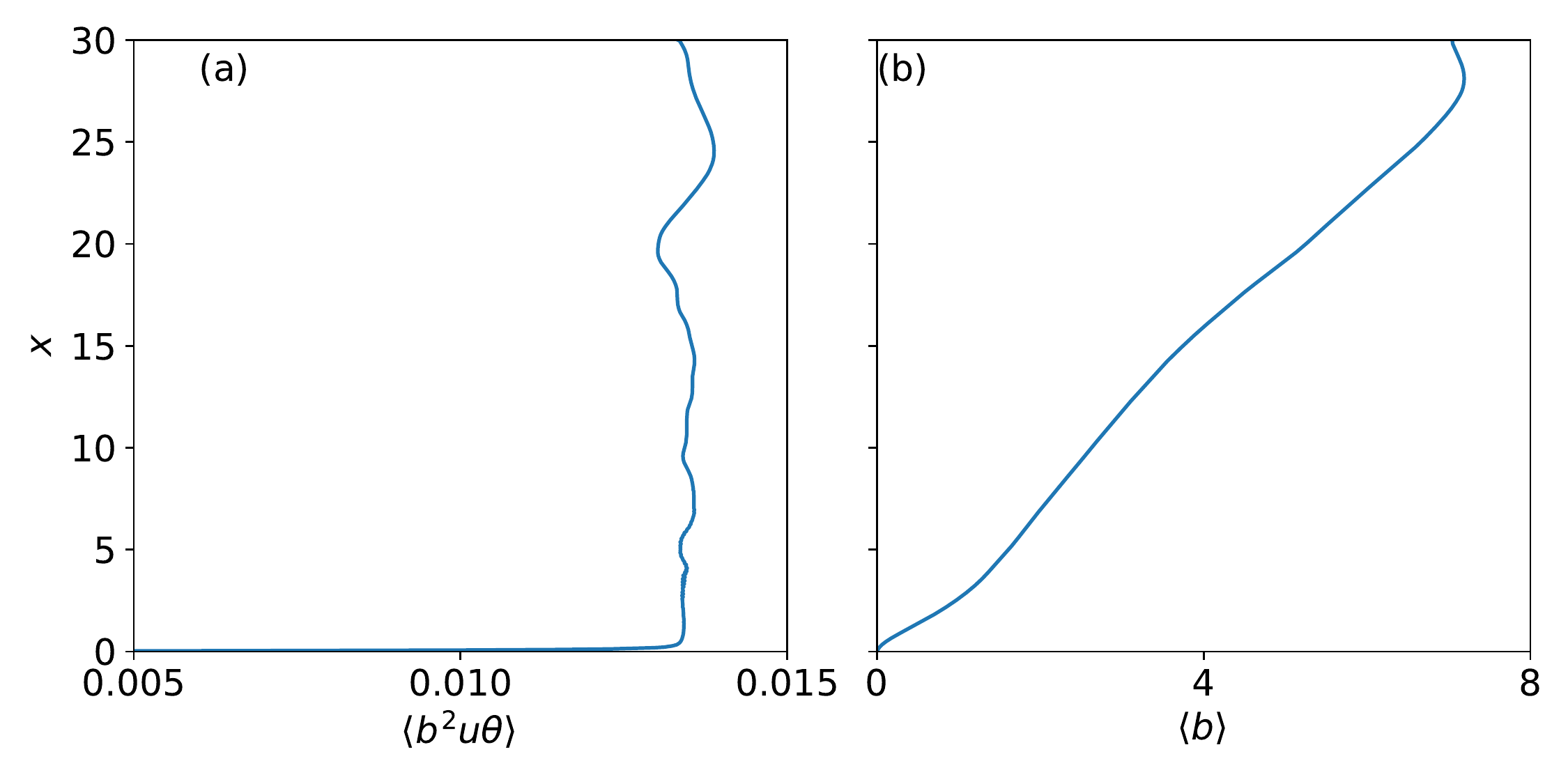} 
\end{center}

\caption{\label{fig:pure_plume_stats} (a): the buoyancy flux $\langle b^2 u\theta \rangle$,
where $\langle \rangle$ denotes an average over the plane and over the time range $200<t<400$, for a plume at Reynolds number $Re=1000$. (b): the plume width, calculated using the mass and momentum fluxes (see eq. \ref{eq:plume_width}).}
\end{figure}

\section{Results and Discussion\label{sec:Results}}

\subsection{Simulations in one dimension \label{subsec:one_d}}

Equilibrium turbulent plumes, like all free-shear flows, entrain ambient fluid at a rate proportional
to a characteristic velocity \citep{MTT1956}. In general, the diffusivities
of temperature and vapour could be different, making the problem more
complex. Since the diffusivities of temperature and vapour (and liquid)
are here assumed to be equal (\S \ref{subsec:thermodynamics}),
a plume that carries vapour in addition to temperature behaves exactly
like a classical plume. This simplification allows a model, following
\cite{MTT1956}, to be written for such a plume. 

Consider a plume of width $b$ and velocity \textbf{$u$}. The mass
and momentum fluxes are $M=b^{2}u$ and $P=b^{2}u^{2}$. The vapour
and liquid fluxes are $F_{v}=b^{2}ur_{v}$ and $F_{l}=b^{2}ur_{l}$.
Assuming a steady state in time, the equations govering the evolution
of these quantities with height are 
\begin{align}
\frac{d}{dx}M & =2\alpha\left|P\right|^{1/2}\nonumber \\
\frac{d}{dx}P & =\frac{M}{P}\left(F+r_{0}\left\{ \left(\chi-1\right)\left(F_{v}-r_{v}^{\infty}M\right)-F_{l}\right\} \right)\nonumber \\
\frac{d}{dx}F & =-M\left(\Gamma_{u}-\Gamma_{0}\right)+L_{2}C_{d}\nonumber \\
\frac{d}{dx}F_{v} & =-C_{d}+2\alpha\left|P\right|^{1/2}r_{v}^{\infty}\nonumber \\
\frac{d}{dx}F_{l} & =C_{d}\label{eq:1D_steady}
\end{align}

In order to avoid the singularity at the origin, equations \ref{eq:1D_steady}
are integrated from initial conditions $M=1,$ $P=1$, $r_{v}=1$,
$\theta=1$, $r_{l}=0$. For a given $\epsilon$, the two parameters
controlling the dynamics are, as before, $\Gamma_{0}$ and $s_{\infty}$.
Note that $\alpha$ is now an externally supplied parameter, and the system of Equations \ref{eq:1D_steady} is thus not closed. In fact, finding the dependence of $\alpha$ on the buoyancy distribution in the flow is an open problem with deep implications for climate modelling \citep{RN2011PNAS}.  We use a
value of $\alpha=0.08$ for the results reported here. 

\begin{figure}
\includegraphics[width=0.49\columnwidth]{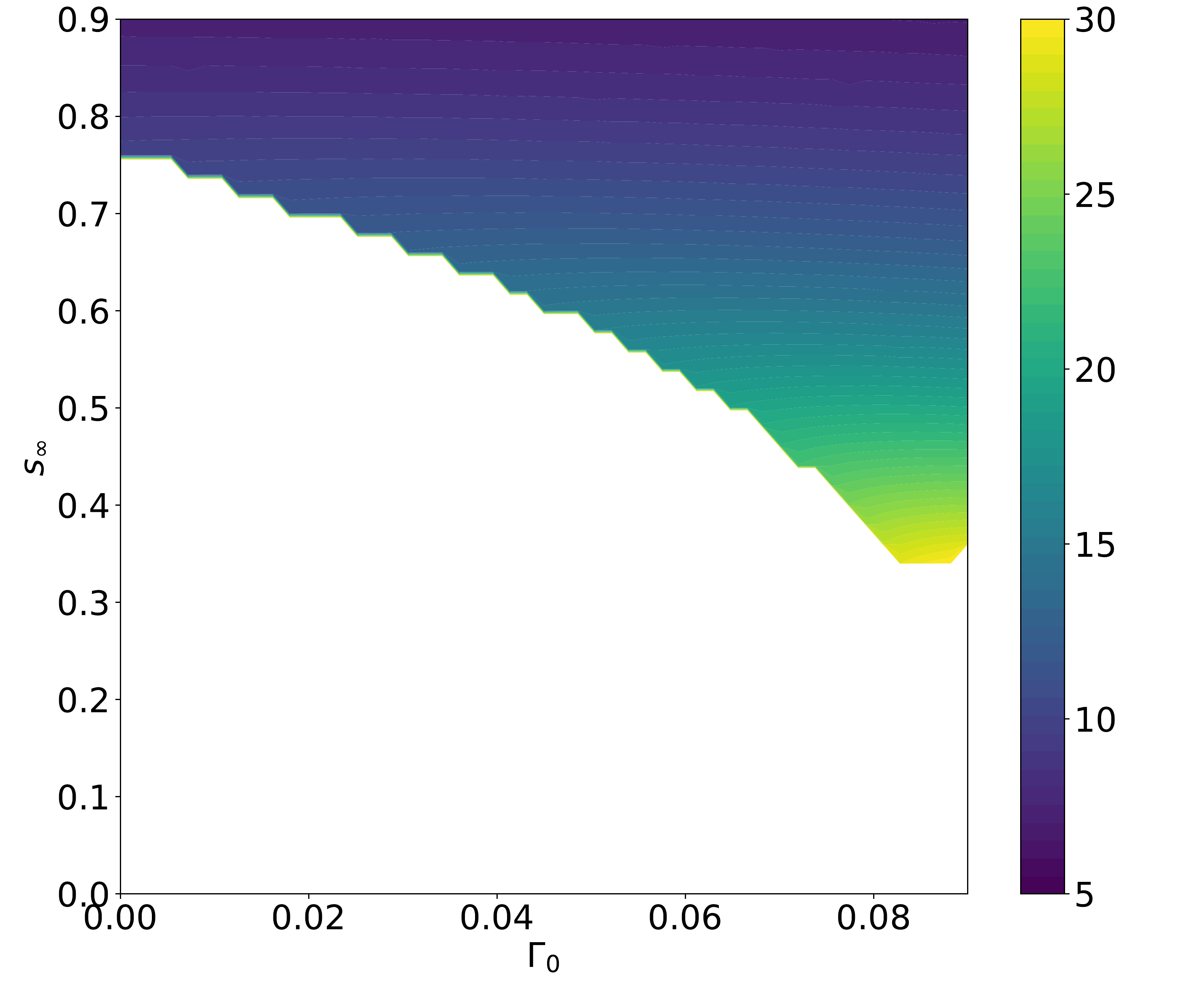}
\includegraphics[width=0.49\columnwidth]{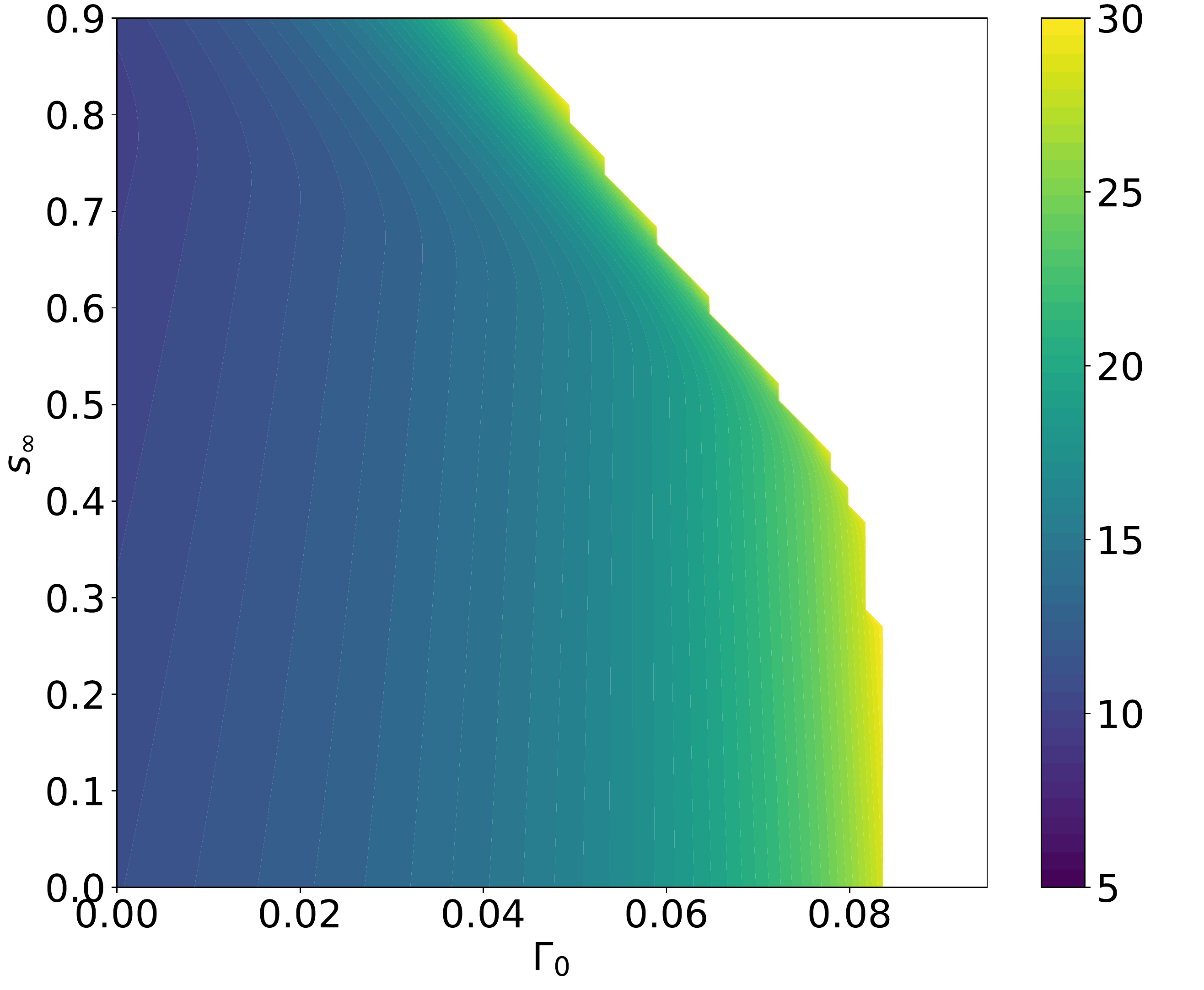}

\caption{\label{fig:LCL_1d} Left: the lifting condensation level predicted
from the 1D model in equations \ref{eq:1D_steady}; and right: 
the maximum height $h_{max}$ reached;  as a function
of $s_{\infty}$ and $\Gamma_{0}$.}
\end{figure}

The evolution of this idealised plume carrying vapour shows a clear
lifting condensation level, defined as the height $x$ where the liquid
flux is greater than zero. This is plotted as a function of $\Gamma_{0}$
in Figure \ref{fig:LCL_1d}. Since a constant value of $\alpha$ is reasonable for turbulent plumes, and since the cloud flow behaves like a turbulent plume until condensation sets in at the lifting condensation line (LCL), the results obtained for the height of the LCL may be considered meaningful.

\subsection{3D simulations}

The benefit of formulating the problem in terms of the nondimensional
parameters $L_{1},L_{2},\Gamma_{d},\epsilon,Re,\Gamma_{0},s_{\infty}$
(\S \ref{subsec:Parameters}) is that the same formulation can
then be applied to any combination of substances (clouds containing
methane vapour and liquid methane in an atmosphere composed mainly
of nitrogen, say). Fixing the combination of fluids (air and water
vapour/liquid for terrestrial clouds) fixes the parameters $L_{1},L_{2},\Gamma_{d}$,
leaving the other four parameters as control parameters. Unless otherwise
stated, we fix the Reynolds number $Re=1000$ and the Atwood number
$\epsilon=1/30$. Thus, the dynamics of the formation of a cloud (which
we will define more precisely in the following) is governed by the
combination of only two properties of the ambient: the relative humidity
of the ambient $s_{\infty}\left(x\right)$, and the lapse rate of
the ambient $\Gamma_{0}$. 

First, we note that until the flow carrying vapour reaches the lifting
condensation level, its dynamics are similar to that of a plume in
a stratified ambient. This can be seen for the case $\Gamma_{0}=0.08$,
$s_{\infty}=0.6$ in Figure \ref{fig:no_cloud_sinf60_Gam8} where no condensation occurs. As we
argued in \S \ref{subsec:stratified_plumes}, the smaller the
lapse rate of the ambient, the more stable the stratification is and
the harder it is for the plume to overcome it and keep rising. Without
the energy input from condensation, therefore, plumes in stably stratified
ambients will die out. Whether condensation can help overcome the
stable stratification depends on if, before the plume has reached
its maximum height, it has either carried up enough vapour with it, or entrained enough
vapour from the ambient. 

From the above discussion, it may be expected that if the stratification
is very stable, a larger background relative humidity will be required
for the plume to be sustained. For ambients which are nearly neutrally
stratified, even reasonably small levels of background relative humidity
will sustain the plume. \\

\begin{figure}
\includegraphics[width=1\columnwidth]{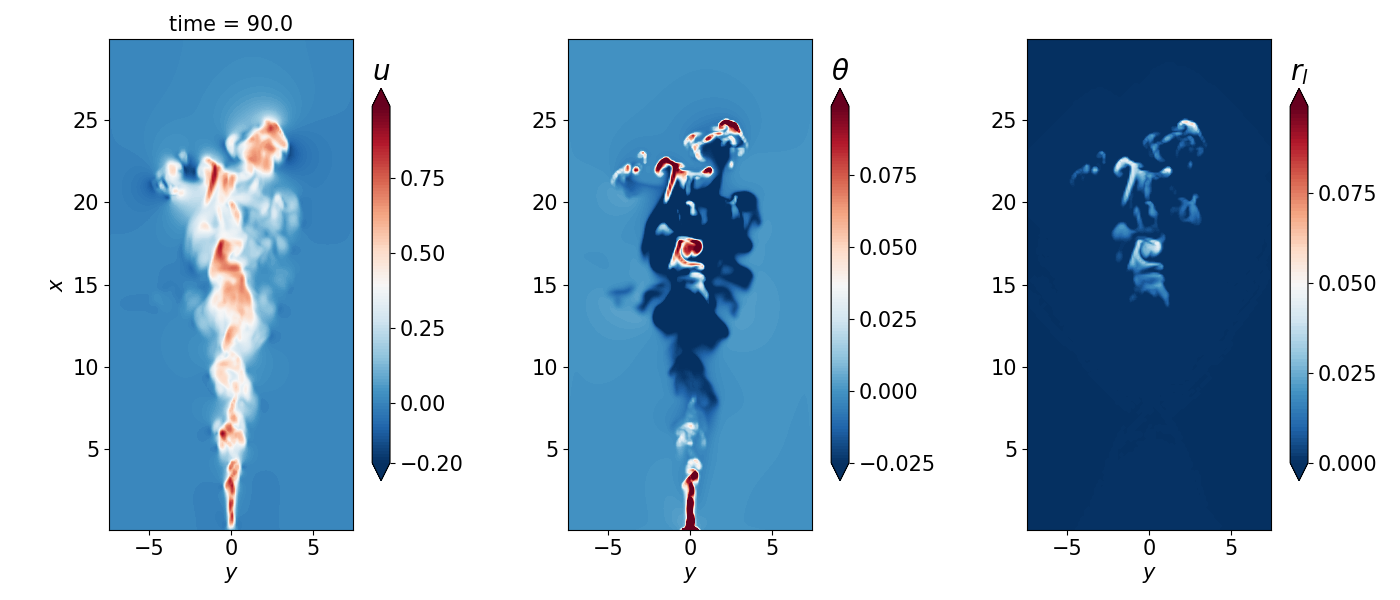}
\includegraphics[width=1\columnwidth]{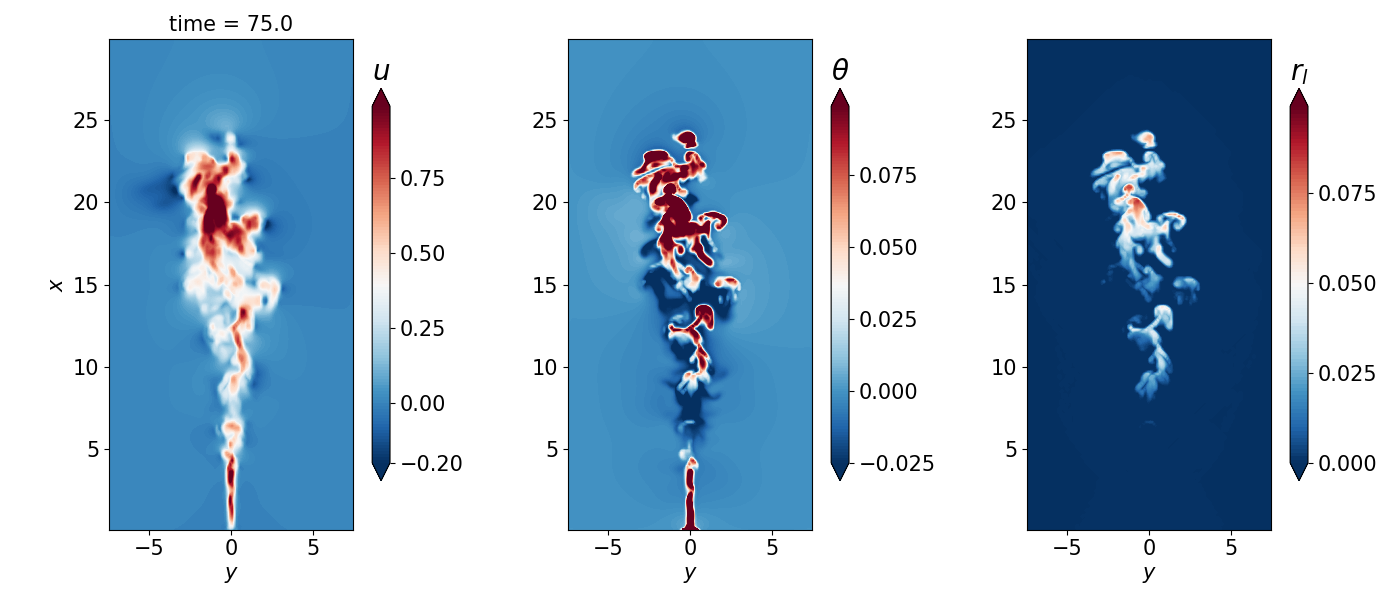}
\caption{\label{fig:cloud_sinf60gam9_sinf80gam8} The velocity, temperature excess, and liquid water mixing ratio for (top) $\Gamma_0=0.09$, $s_{\infty}=0.6$; and (bottom) $\Gamma_0=0.08$, $s_{\infty}=0.8$. The LCL is at a much lower altitude and the ensuing convection is much more vigorous in the latter than in the former. The contours are plotted on the $z=0$ plane.}
\end{figure}

\begin{figure}
\includegraphics[width=1\columnwidth]{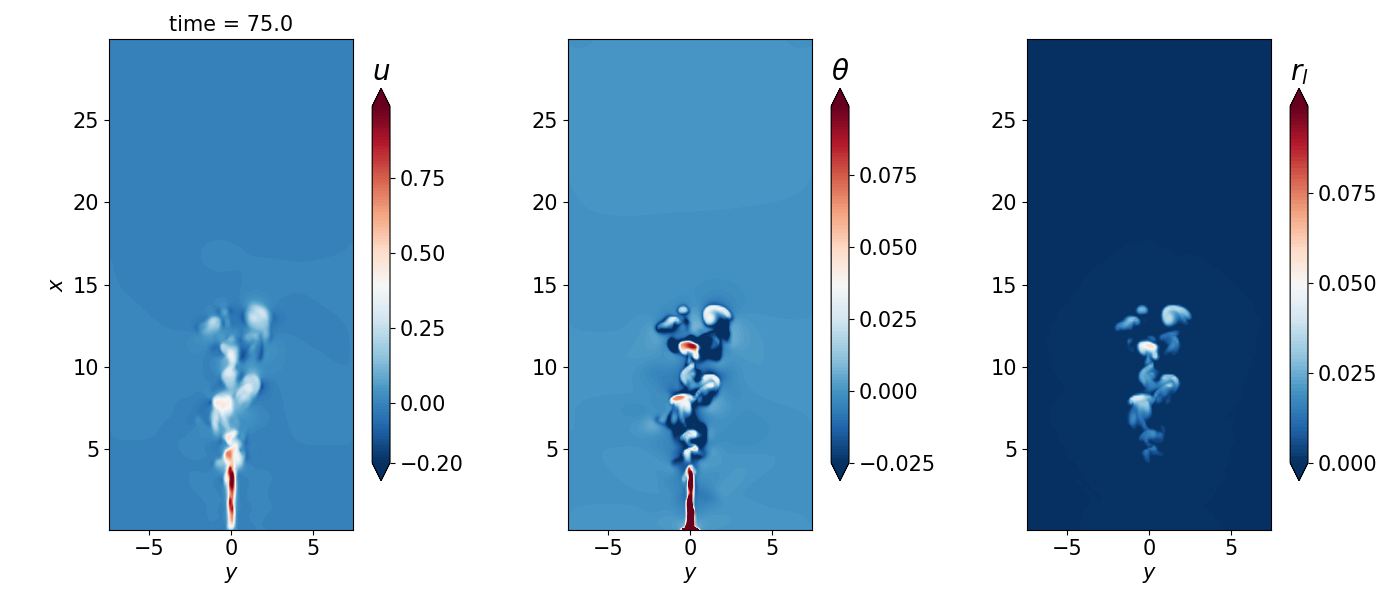}
\includegraphics[width=1\columnwidth]{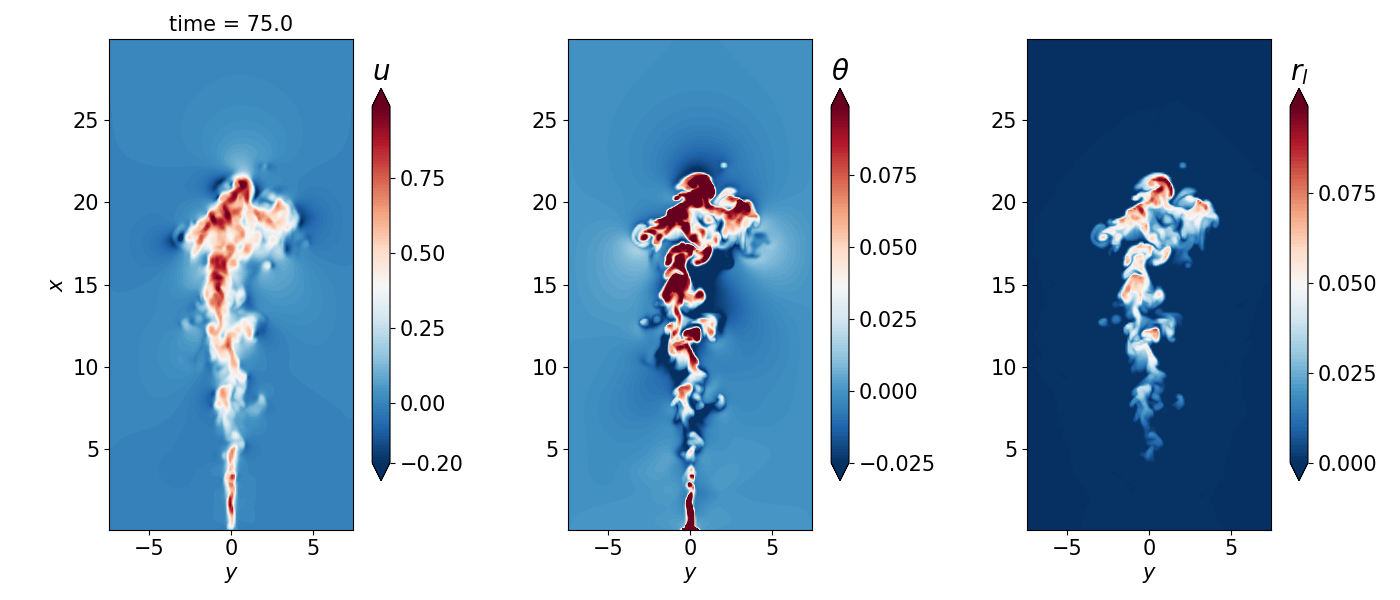}
\caption{\label{fig:cloud_sinf90gam6_sinf90gam6d5} Velocity, temperature 
and liquid water mixing ratio contours for (top) $\Gamma_0=0.06$; and (bottom) $\Gamma_0=0.065$, with $s_{\infty}=0.9$ in both cases. A small change in the lapse rate affects the dynamics. The contours are plotted on the $z=0$ plane as before.}
\end{figure}

Visible clouds are those that contain liquid water droplets. 
In Figure \ref{fig:cloud_sinf60gam9_sinf80gam8}, we show snapshots of the
evolution for two cases from table \ref{tab:phase_plane}. In the first ($s_\infty = 0.6, \Gamma_0 = 0.09$),
the cloud that forms has an LCL at $x\approx 15$, and is fragmented. The second case 
($s_\infty = 0.8, \Gamma_0 = 0.08$) shows a much more vigorous cloud. These cases may be compared with
the case presented in Figure \ref{fig:no_cloud_sinf60_Gam8}, where no cloud forms.
Figure \ref{fig:cloud_sinf90gam6_sinf90gam6d5} shows cloud formation in a cases where the atmosphere
is almost saturated, with $s_\infty=0.9$, and $\Gamma_0 = 0.06$ and $0.065$. These values
of $\Gamma_0$ are close to the boundary of cloud formation. As a result, the small increase in 
$\Gamma_0$ leads to a dramatic increase in the vigorousness of the convection.

\subsection{Vapour Boundary Conditions \label{subsec:vapour_bcs}}
\begin{figure}
\includegraphics[width=1\columnwidth]{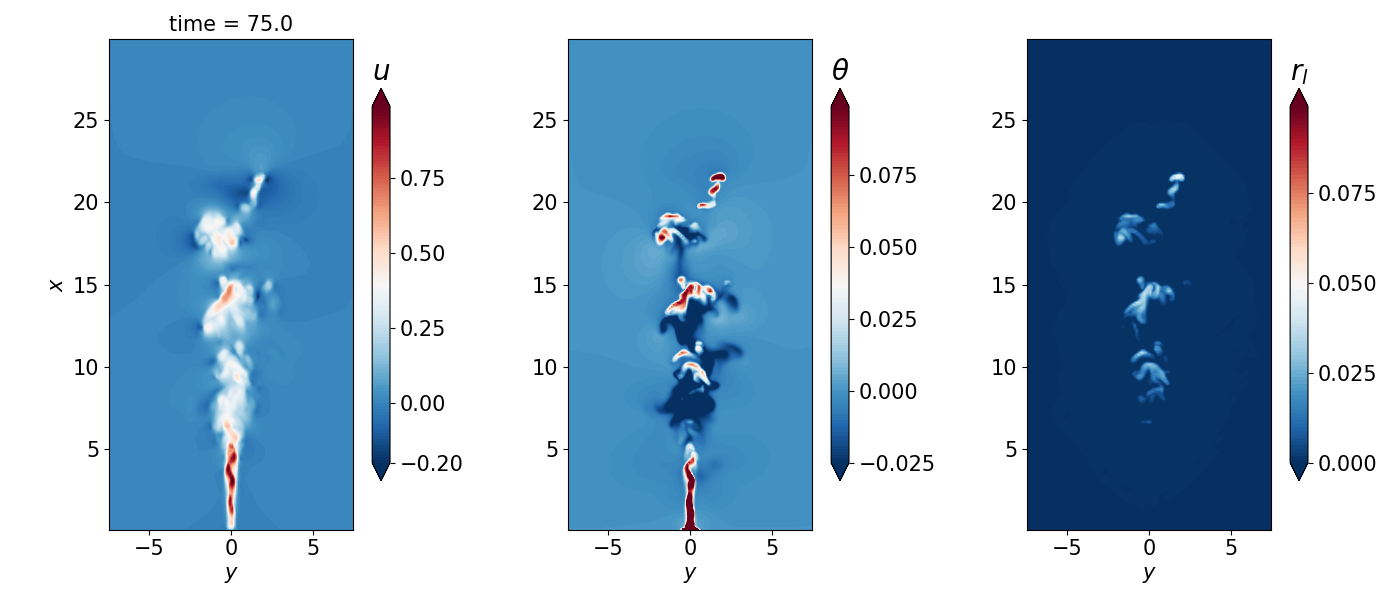}
\includegraphics[width=1\columnwidth]{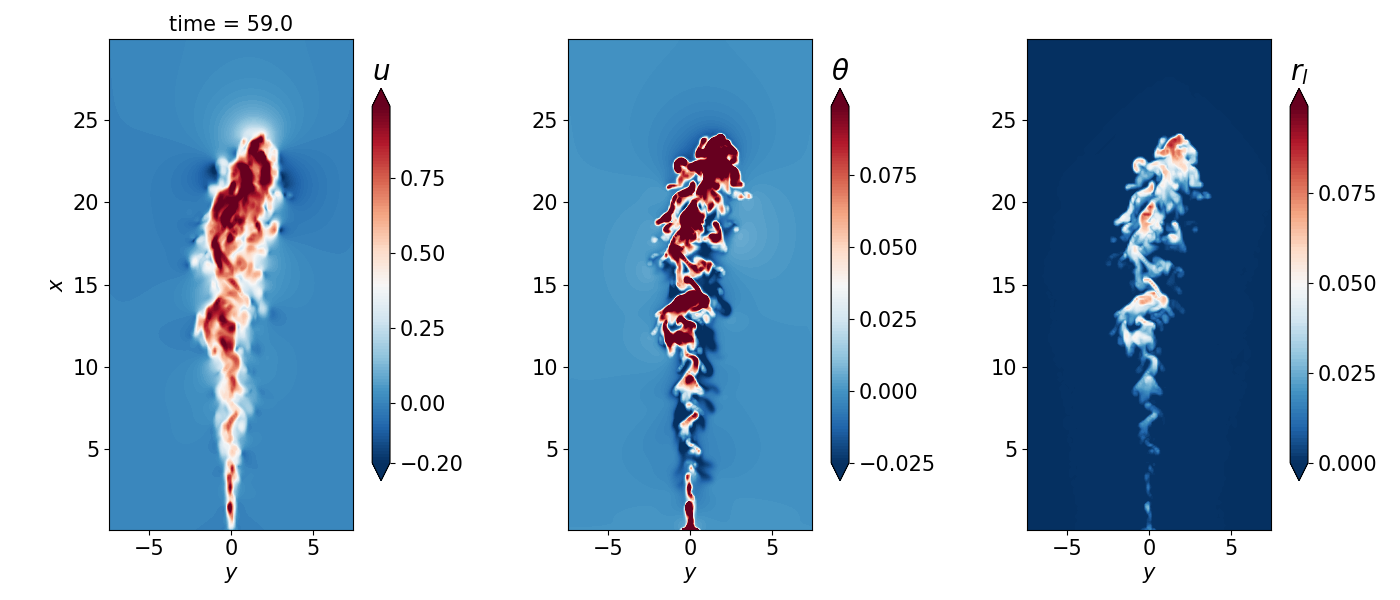}
\caption{\label{fig:basesat_sinf70_gam9} Iso-contours on the $z=0$ plane for $\Gamma_0=0.08$ and $s_{\infty}=0.8$, with (top) $r_v = s_\infty r_s\left(\theta=0\right)$;  and (bottom) $r_v = r_s\left(\theta=1\right)$  at the hot patch. The contours are plotted on the $z=0$ plane as before. Note the presence of liquid for all heights in the latter case. Convection becomes increasingly more vigorous as more vapour is introduced into the flow through the hot patch. These may be compared with Fig. \ref{fig:cloud_sinf60gam9_sinf80gam8}(b)}
\end{figure}

As mentioned in \S \ref{subsec:Boundary-Conditions}, we study three kinds
of boundary conditions for the vapour. These boundary conditions control the
amount of vapour introduced into the plume through the hot patch. With increasing
vapour in the flow, condensation occurs sooner, thereby lowering the lifting condensation
level. In fact, for boundary conditions of the second type (vapour saturated at $\theta=1$),
the lifting condensation level could be at $x=0$. These changes, however, do not change
the behaviour of the flow qualitatively. A comparison between these boundary conditions
is shown in Figure \ref{fig:basesat_sinf70_gam9}.

\noindent \begin{center}
\begin{figure}
\noindent \begin{centering}
\includegraphics[width=0.8\columnwidth]{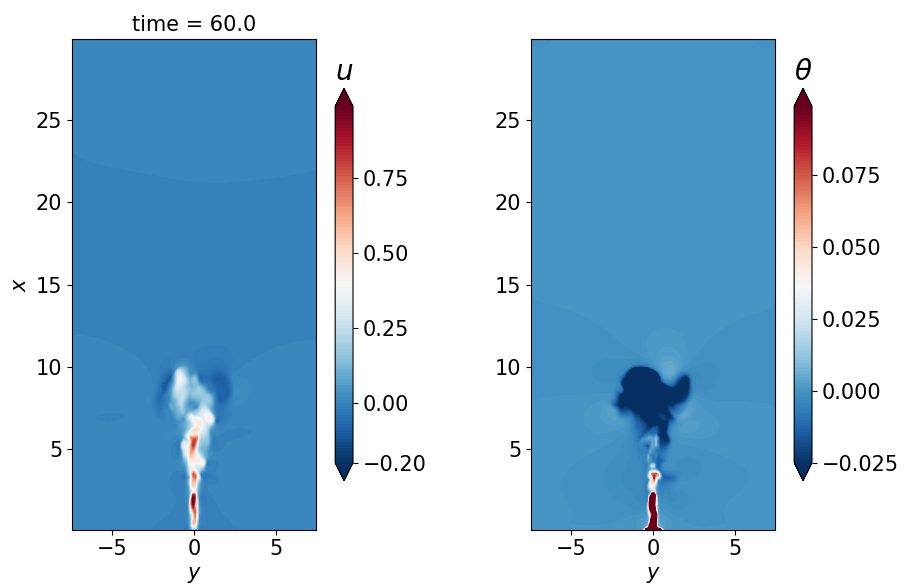}
\par\end{centering}
\caption{\label{fig:no_cloud_sinf60_Gam8}The velocity and temperature for
a case where no cloud forms (i.e. the liquid water content $r_{l}=0$).
The lapse rate is $\Gamma_0=0.08$. The relative humidity
$s_{\infty}=60\%$. At the time-instant shown $\left(t=60\right)$,
the plume has already hit the maximum height $H_{T}$ and the head
of the plume has descended back towards its equilibrium height. Note
that $\theta<0$ at the plume-top.}
\end{figure}
\par\end{center}

\subsection{Cloud--no-cloud boundary \label{subsec:cloud_nocloud_boundary}}

\begin{figure}
\includegraphics[width=0.5\columnwidth]{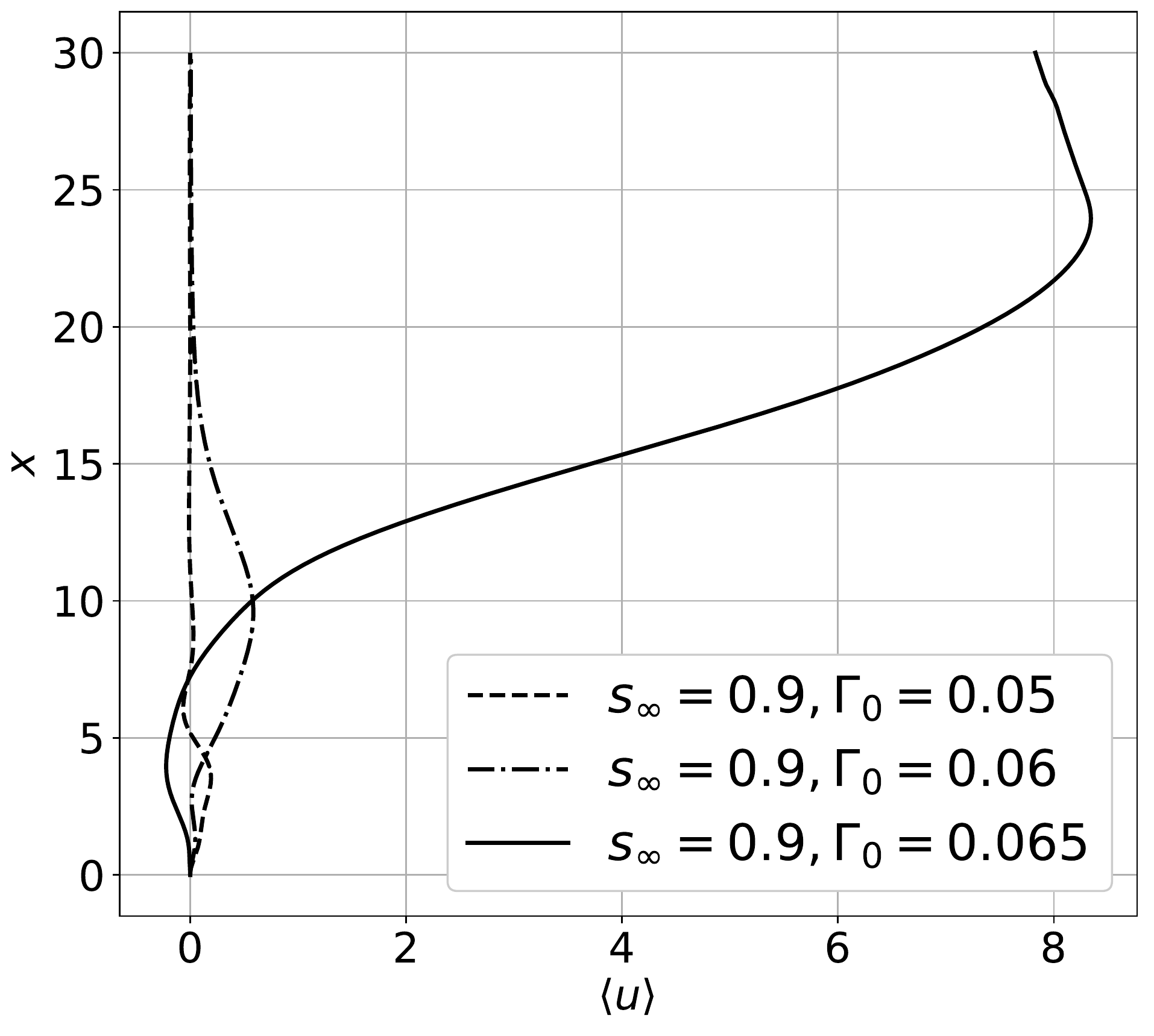}\includegraphics[width=0.55\columnwidth]{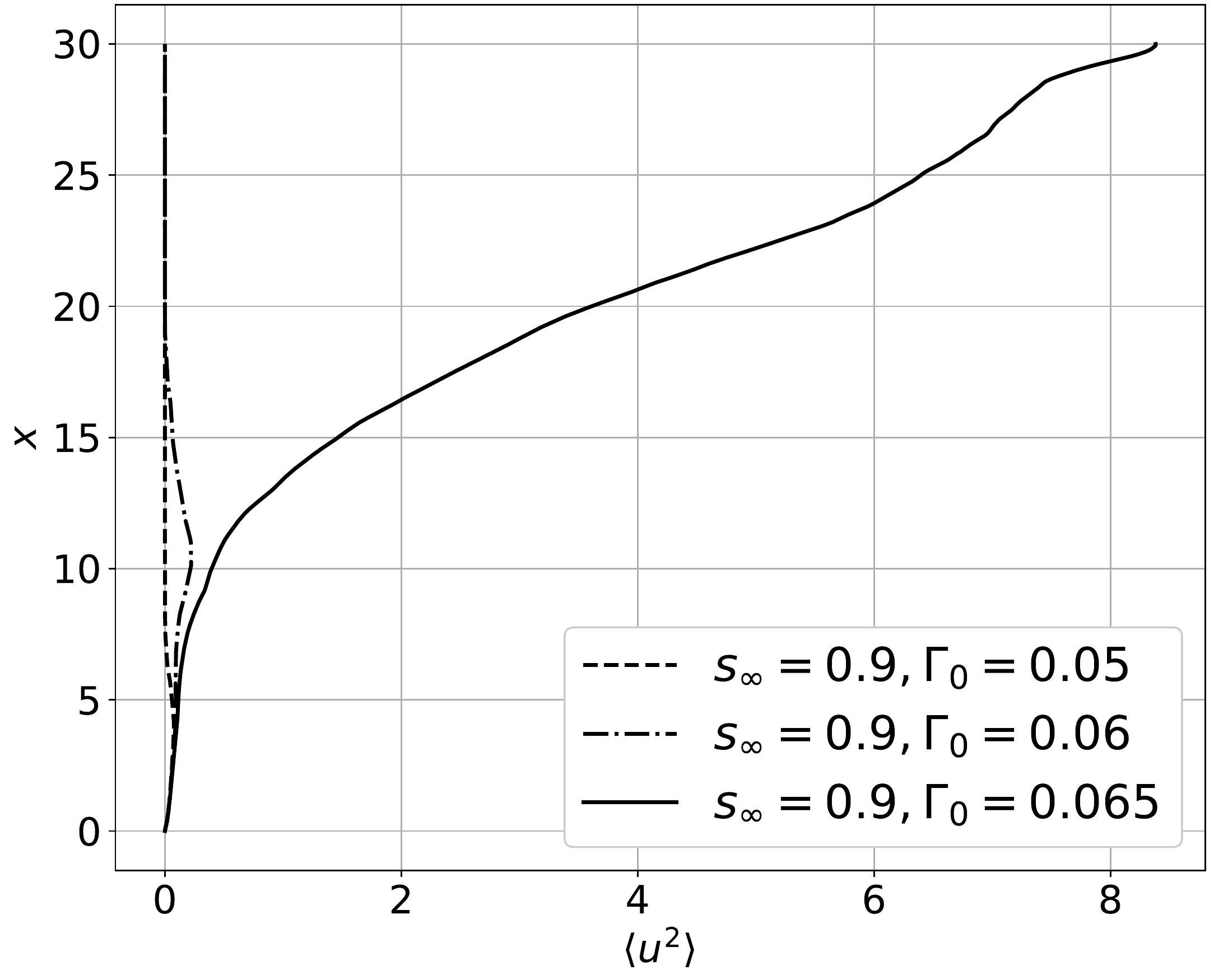}
\includegraphics[width=0.5\columnwidth]{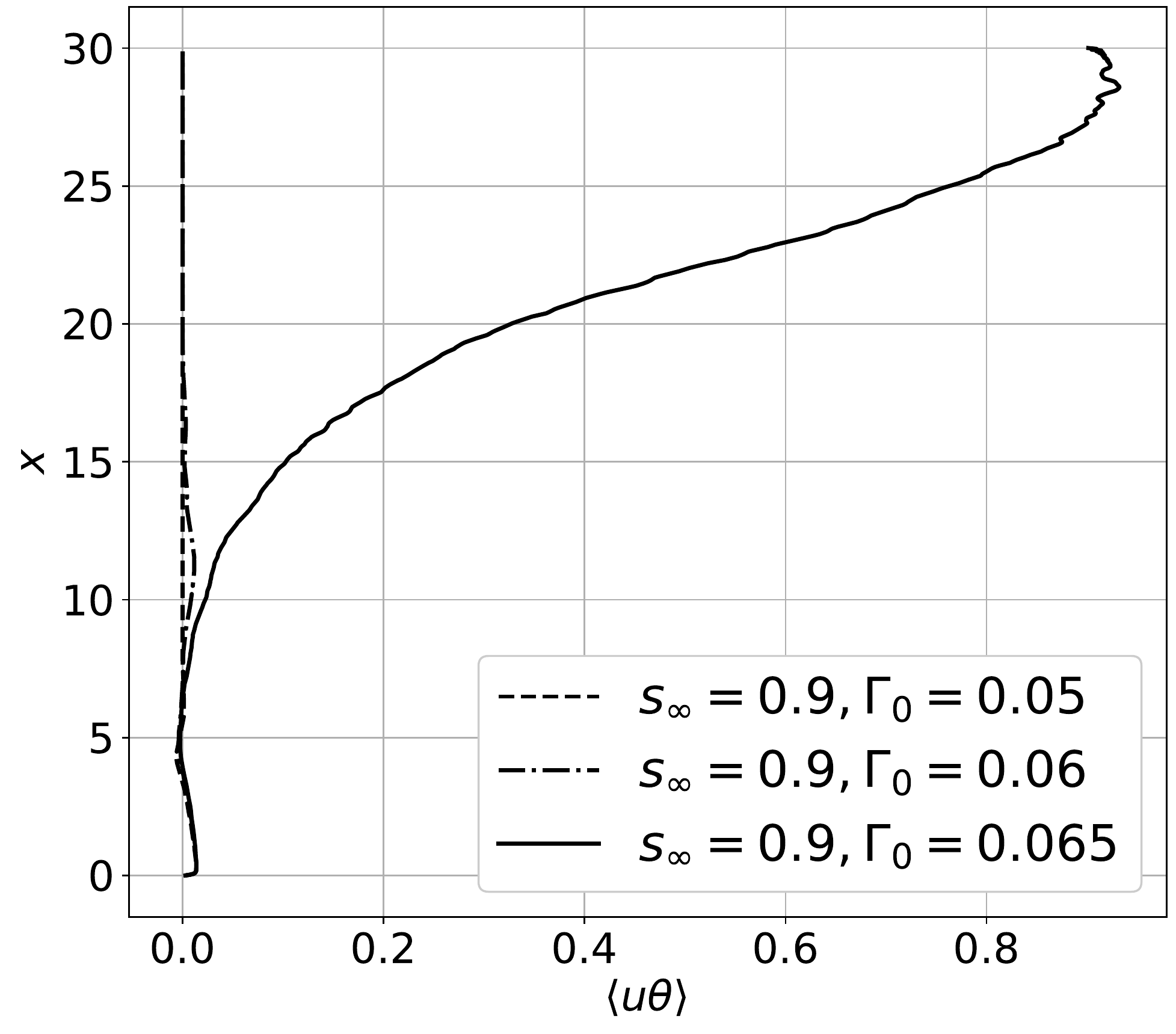}\includegraphics[width=0.5\columnwidth]{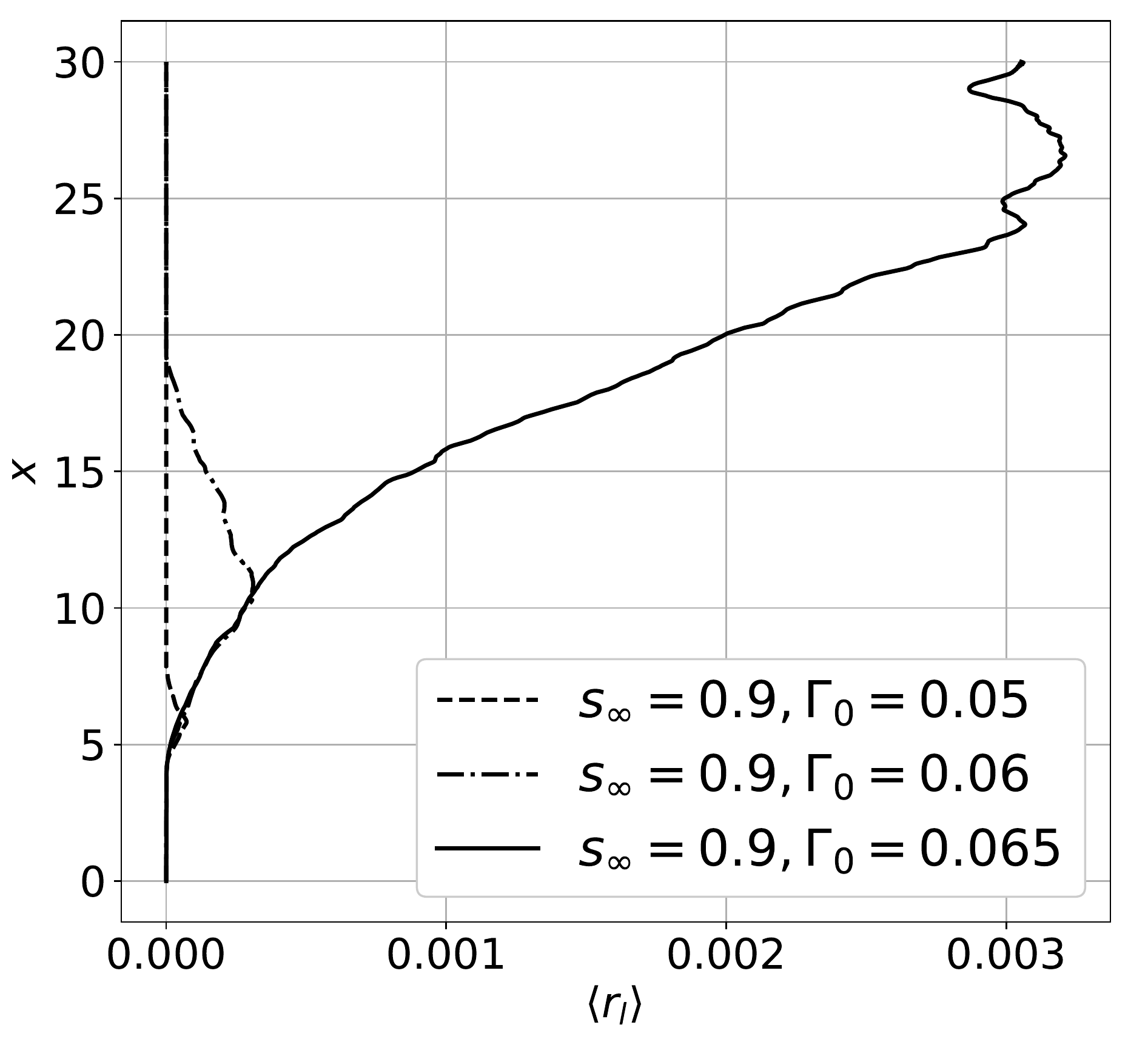}
\caption{Mass (top-left), momentum (top-right) and buoyancy (bottom-left) fluxes, and the average liquid water content ($r_l$) averaged from $t=50$ to $t=100$ as a function of height, showing the influence of increasing lapse rate. The ambient vapour content is fixed at $s_\infty = 0.9$.
\label{fig:mass_mom_buoy_vs_gam}}
\end{figure}

\begin{figure}
\includegraphics[width=0.5\columnwidth]{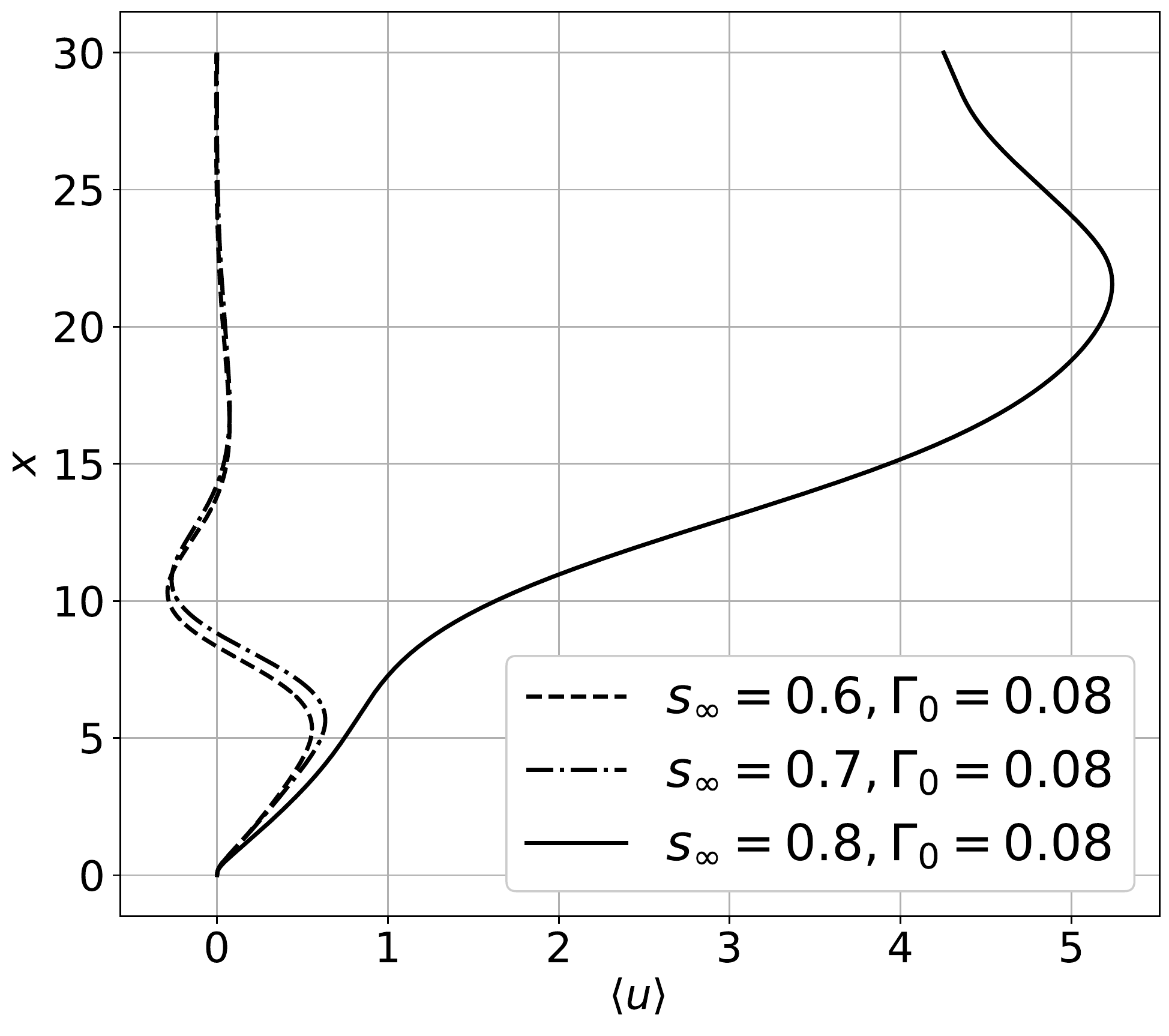}\includegraphics[width=0.5\columnwidth]{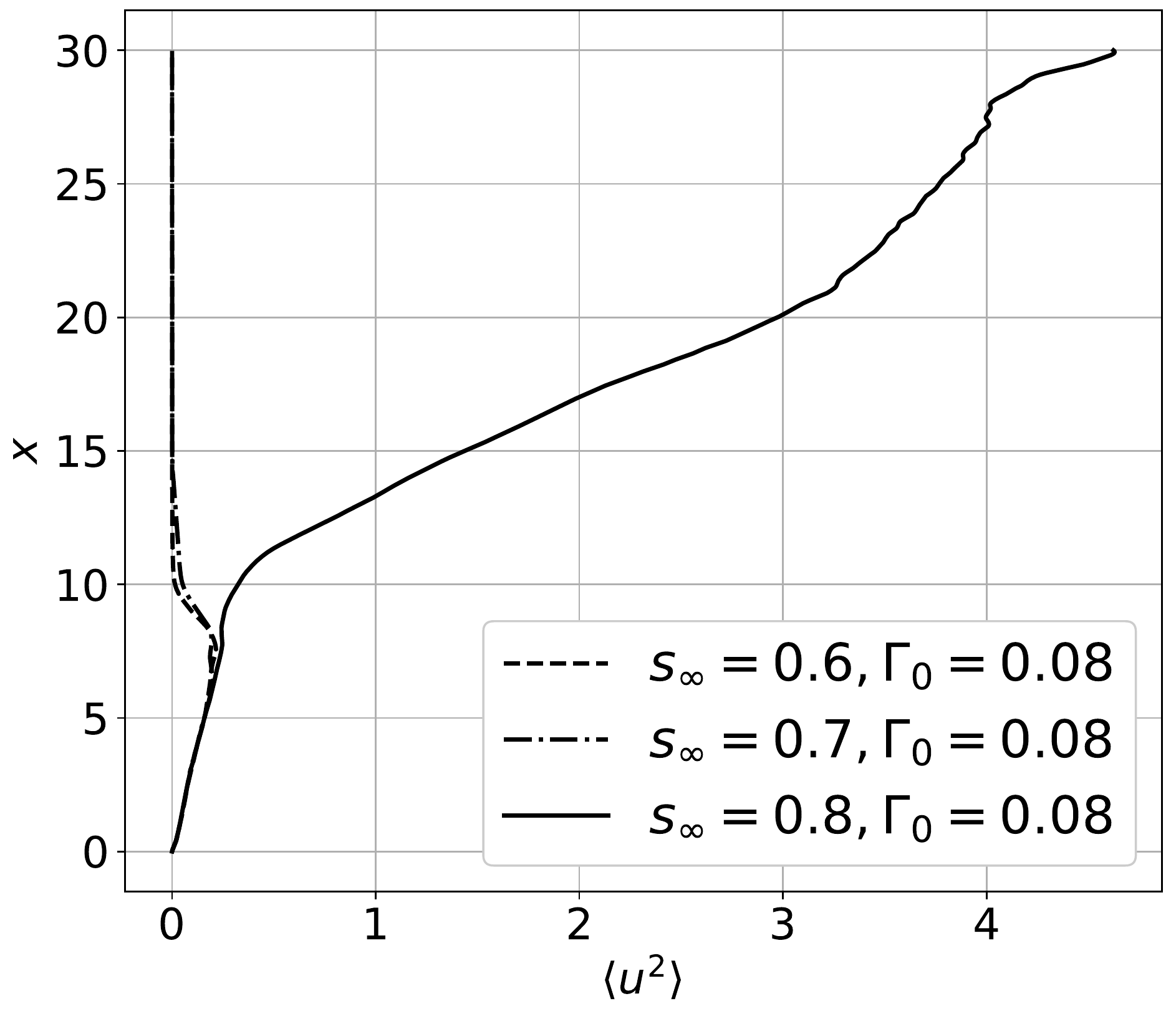}
\includegraphics[width=0.5\columnwidth]{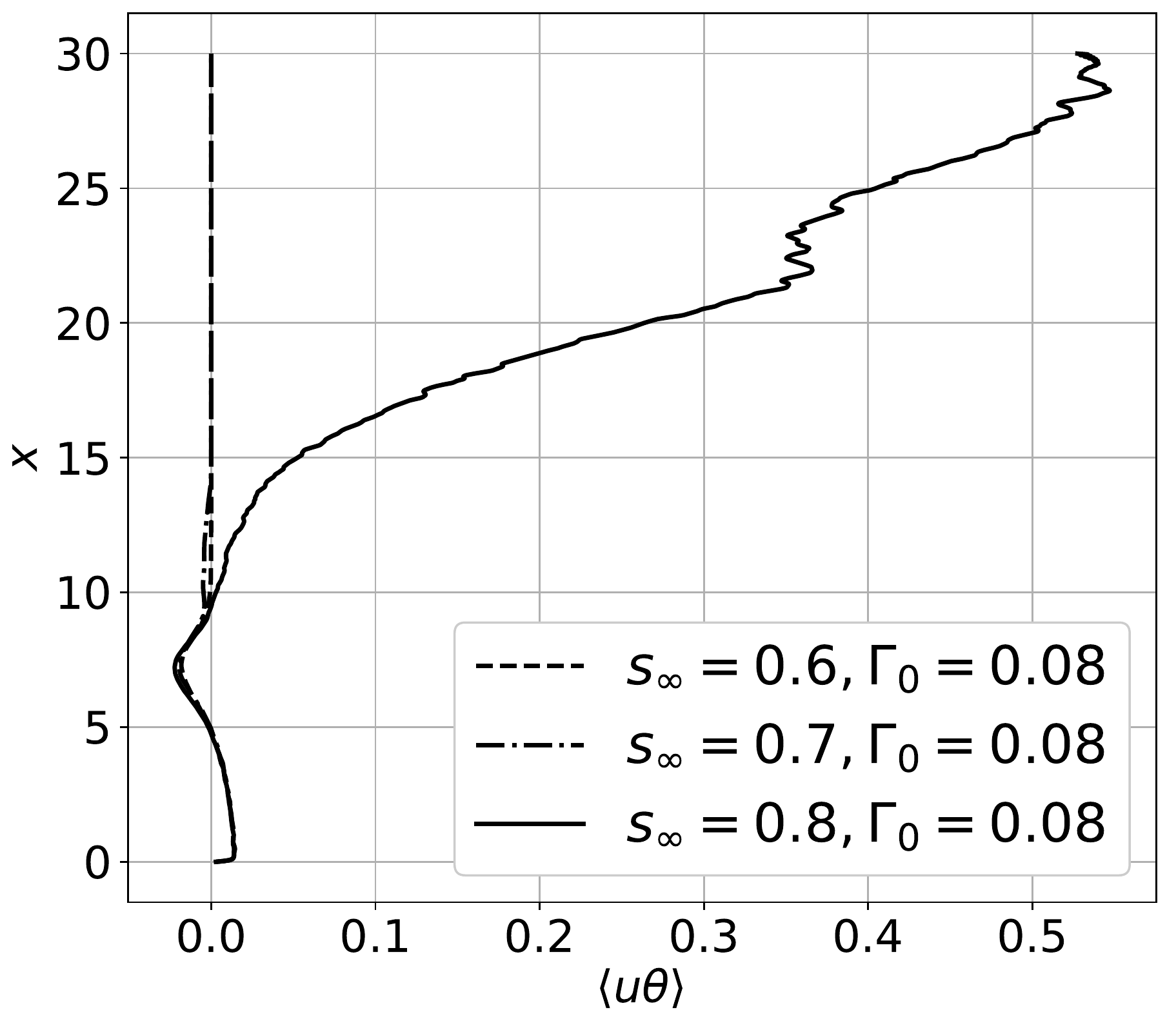}\includegraphics[width=0.52\columnwidth]{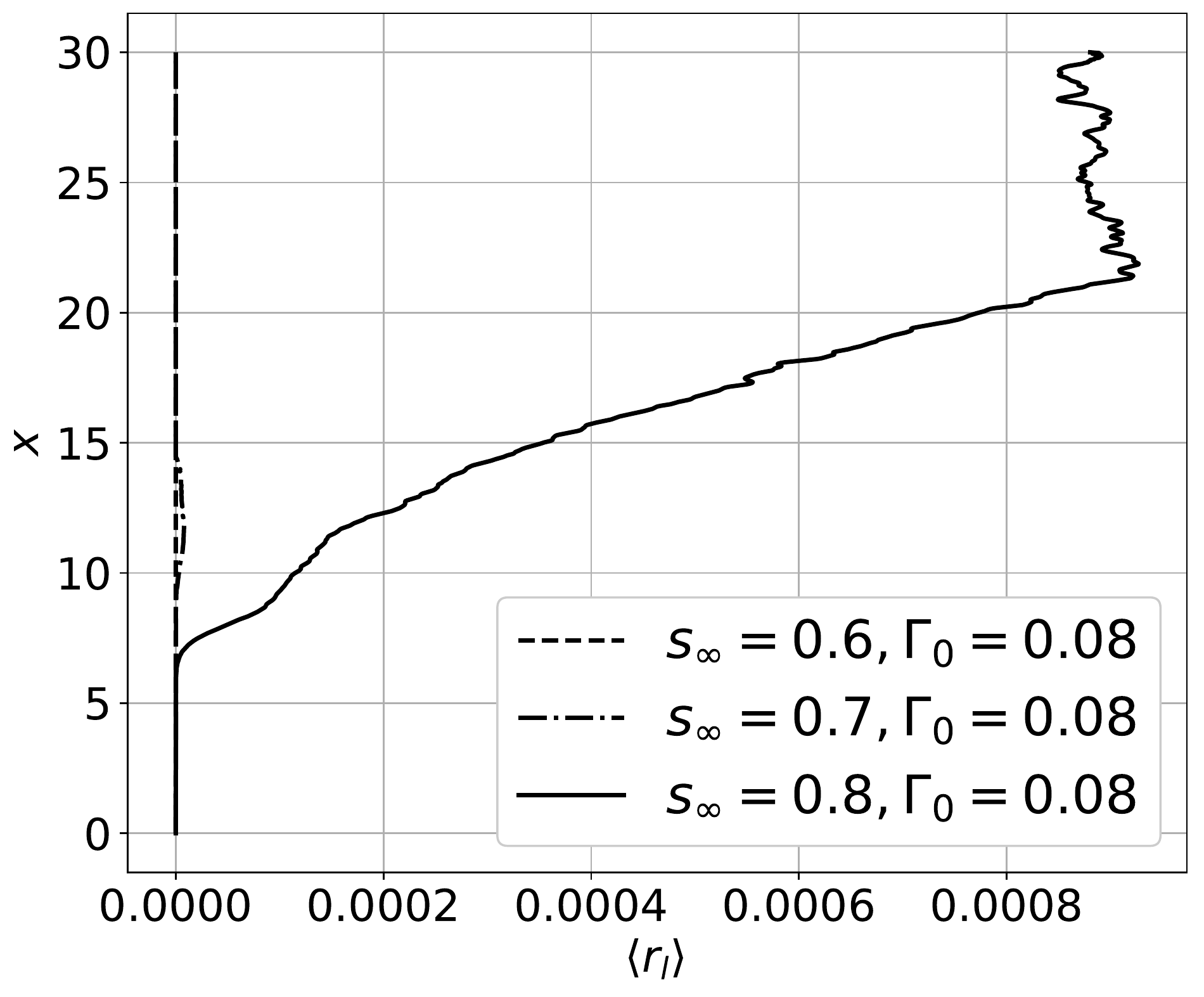}
\caption{ As in Fig. \ref{fig:mass_mom_buoy_vs_gam}, but with a fixed lapse rate and showing the effects of varying $s_\infty$.  \label{fig:mass_mom_buoy_vs_sinf}}
\end{figure}

We now quantify the behaviour of moist plumes as a function of $s_{\infty}$
and $\Gamma_{0}$. For the cases presented in Figs. \ref{fig:no_cloud_sinf60_Gam8}-\ref{fig:cloud_sinf90gam6_sinf90gam6d5},
Figures \ref{fig:mass_mom_buoy_vs_gam} and \ref{fig:mass_mom_buoy_vs_sinf} show the fluxes of mass, momentum and buoancy as a function of height. The cases where clouds form (i.e. where there is a finite amount of liquid water present; also shown in Figs. \ref{fig:mass_mom_buoy_vs_gam} and \ref{fig:mass_mom_buoy_vs_sinf}) show a dramatic increase in these fluxes by up to two orders of magnitude. As expected, these fluxes increase with increasing $s_\infty$ and $\Gamma_0$. Note that the fluxes with higher ambient vapour content are much larger, even when clouds form in both cases, showing the crucial role played by the energy provided by condensation. 

The balance between the background relative
humidity and the lapse rate depicted in table \ref{tab:phase_plane}.
In this table, which is drawn in the phase-space of $s_{\infty}$
and $\Gamma_{0}$, the origin (at top left) is an ambient that is
completely dry and is very stably stratified. Thus, no cloud will
ever form at the origin. The further away from the origin we venture
in this space, the easier it is for clouds to form (though we note
that traversing along one axis may not be as easy physically as traversing
along the other). 

\begin{figure}
\noindent \begin{centering}
\includegraphics[width=0.6\columnwidth]{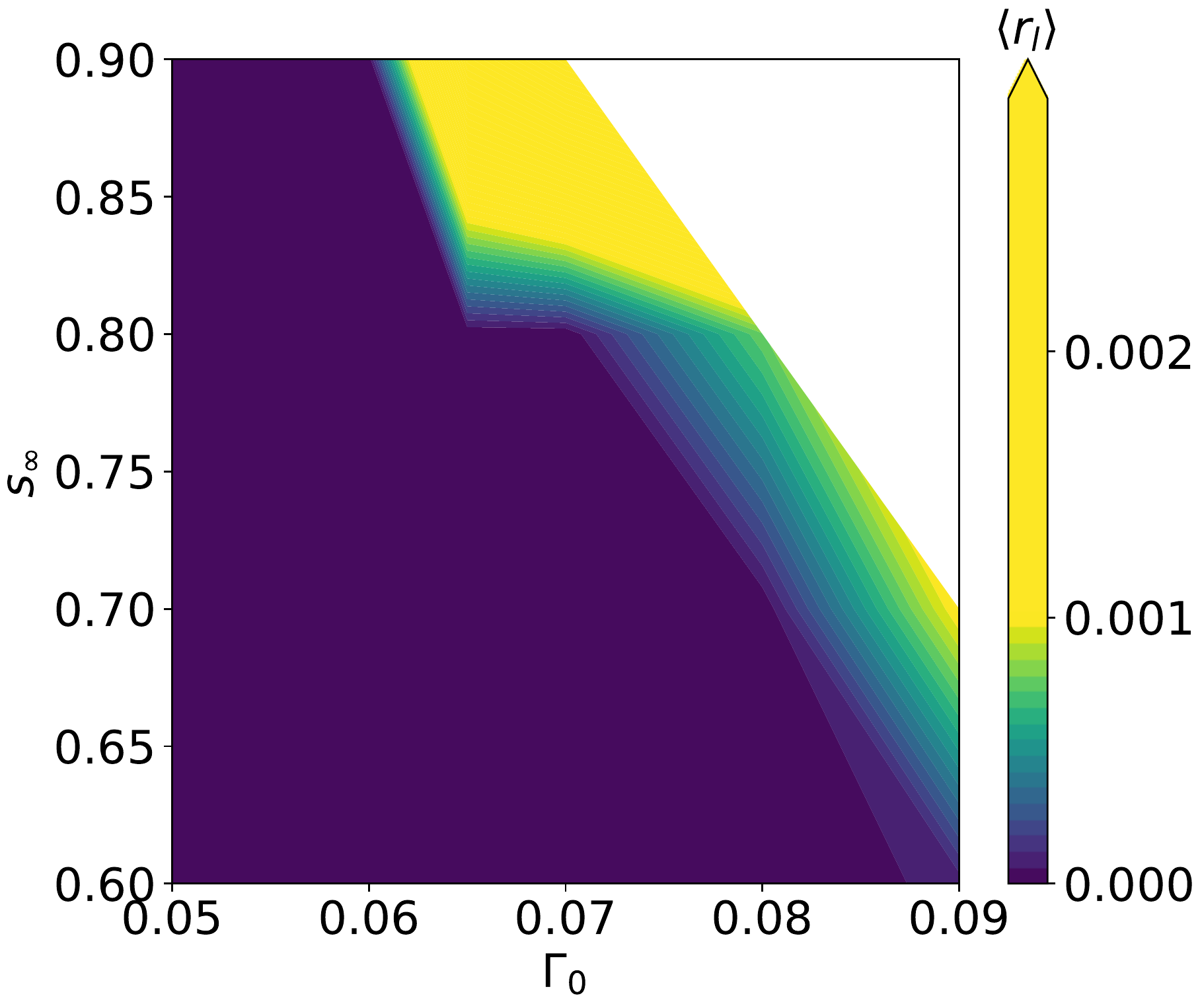}
\par \end{centering}
\caption{\label{fig:cloud_nocloud} The boundary separating the region
in the $s_{\infty}$-$\Gamma_{0}$ space where clouds can or cannot
form, drawn using the average liquid flux in the upper half of the domain.}
\end{figure}

This balance is further quantified in Figure \ref{fig:cloud_nocloud}, where the average liquid water content in the upper half of the domain ($15 < x < 30$) over $50<t<100$ is plotted as a function of $s_\infty$ and $\Gamma_0$. This plot and Table \ref{tab:phase_plane} may be compared with the predictions from the 1D model in Figs. \ref{fig:LCL_1d}.

%
%

\begin{table}
\noindent \begin{centering}
\begin{tabular}{|c|c|c|c|c|c|c|}
\hline 
$s_{\infty}$ \textbackslash{}$\Gamma_{0}$ & $0.05$ & $0.06$ & $0.065$ & $0.07$ & $0.08$ & $0.09$\tabularnewline
\hline 
$0.6$ & x & x & x & x & $\otimes$ & $\odot$  \tabularnewline
\hline 
$0.7$ & x & x & x & x & $\otimes$ & $\odot$ \tabularnewline
\hline 
$0.8$ & x & x & x & $\otimes$ & $\odot$ & x  \tabularnewline
\hline 
$0.9$ & $\otimes$ & $\odot$ & $\odot$ & $\odot$ & x & x  \tabularnewline
\hline 
\end{tabular}\hfill{}
\par \end{centering}
\caption{\label{tab:phase_plane}Does a given combination of $s_{\infty}$
and $\Gamma_{0}$ lead to a `cloud'? Legend: `x': no simulation; $\otimes$:
no cloud; $\odot$: cloud. The presence of a ``cloud'' is defined as in Fig. \ref{fig:cloud_nocloud}, by considering the amount of liquid present in the upper half of the domain}
\end{table}


\subsection{Plumes versus thermals: models for cumulus convection \label{subsec:Plumes-vs-thermals}}

As mentioned in \S \ref{sec:Introduction}, cumulus clouds have
been alternatively modelled as steady plumes or unsteady thermals.
The results here suggest that cumulus clouds are best thought of as
steady plumes until the lifting condensation level, and transient
diabatic coherent structures after condensation begins. The nature
of these coherent structures and the entrainment into them remains
a problem of active study.

\pagebreak

\section{Conclusion\label{sec:Conclusion}}

We have presented a framework using which the dynamics of cumulus
clouds can be studied with direct numerical simulations. This framework
relies on the simplification of the Clausius-Clapeyron equation, 
with consistent approximation of both the dynamics and the
thermodynamics, such that the system of equations does not explicitly
contain the absolute temperature or the properties of the fluid.
We use this framework to show that the formation of
cumulus clouds can be described in terms of a small number of parameters,
and that the dynamics depends crucially on the relative humidity and
the lapse rate of the ambient. In terms of the latter pair of parameters,
we show when cumulus clouds may form in a stratified ambient out of
the plumes that emanate from a hot patch. We show that there exists
a boundary in the $s_{\infty}-\Gamma_{0}$ plane which separates the
regions where cumulus clouds can and cannot form. The existence of
this boundary, we show, is robust to changes in $Re$ and $\epsilon$
, although where the boundary appears in the $s_{\infty}-\Gamma_{0}$
phase plane does depend on $Re$ and $\epsilon$.

We posit that the long-standing puzzle of anomalous entrainment in
cumulus clouds can be resolved through direct numerical simulations
of these flows. In this pursuit, our formulation provides a simple
framework which includes all the relevant physics while retaining
the simplicity that enables the use of high-performance computing.
The entrainment in cumulus clouds is the subject of an ongoing study
in the group.

\section*{Acknowledgements}
We wish to thank Vybhav G. Rao, Samrat Rao, Sachin Shinde, Maruthi
NH, Kishore Patel, and Professors Garry Brown, S.M. Deshpande and
Rama Govindarajan for helpful discussions, and Prasanth Prabhakaran,
who wrote the earlier version (\emph{Megha}-4) of the code used here.

\bibliography{references}
\end{document}